\documentclass{ws-ijmpe}
\usepackage[super,compress]{cite}
\begin{document}

\newcommand{\UNIT}[1]{\mbox{$\,{\rm #1}$}}
\newcommand{\MeV}{\UNIT{MeV}}
\newcommand{\GeV}{\UNIT{GeV}}
\newcommand{\GeVc}{\UNIT{GeV/c}}
\newcommand{\TeV}{\UNIT{TeV}}
\newcommand{\AMeV}{\UNIT{AMeV}}
\newcommand{\AGeV}{\UNIT{AGeV}}
\newcommand{\ATeV}{\UNIT{ATeV}}
\newcommand{\fm}{\UNIT{fm}}
\newcommand{\mb}{\UNIT{mb}}
\newcommand{\mub}{\UNIT{\mu b}}
\newcommand{\nb}{\UNIT{nb}}
\newcommand{\fmc}{\UNIT{fm/c}}
\newcommand{\proz}{\UNIT{\%}}
\newcommand{\panda}{$\overline{\mbox P}$ANDA~}
\newcommand{\p}{\partial}

\markboth{Theodoros Gaitanos}{Recent progress on superstrange dynamics}

\catchline{}{}{}{}{}

\title{Recent progress on superstrange dynamics}

\author{Theodoros Gaitanos}

\address{Faculty of Science, School of Physics, 
Aristotle University of Thessaloniki\\
55124 Thessaloniki, Greece\\
tgaitano@auth.gr}

\maketitle

\begin{history}
\received{Day Month Year}
\revised{Day Month Year}
\end{history}

\begin{abstract}
We review the present activities related to hypernuclear production in 
hadronic in-medium reactions at intermediate energies. The status of 
theoretical predictions and experimental evidences for the in-medium 
formation of bound superstrange matter is discussed. Heavy-ion 
collisions and antiproton-induced reactions at energies close to 
strangeness production thresholds create the conditions of hypermatter 
formation. This allows to understand better in-medium hyperon interactions 
and sets constraints on the strangeness sector of the nuclear equation of 
state. 
\end{abstract}


\ccode{PACS numbers: 25.40.Ve, 25.43.+t, 25.75.Dw, 25.80.Pw, 21.65.Mn, 21.80.+a}


\section{Introduction}
The equation of state (EoS) of strongly interacting matter has been one of the most 
researched subject in nuclear and hadron physics over the last $50$ years. 
Concrete investigations on the nuclear EoS were initiated in the early 1970s 
with the realization of the first relativistic nuclear beams 
at BEVELAC in Berkeley~\cite{first1,first2,first3}, followed by more 
selective heavy-ion experiments in centrality, particle identification and 
phase space reconstruction~\cite{ref7,ref9,ref10,fopi1,fopi2,fopirev1,fopirev2}. 
These measurements gave first hints on collective particle 
flow~\cite{ref12,ref6,ref13,ref15,ref16}. 
They allowed first interpretations 
on the nuclear EoS at different density regions beyond 
saturation~\cite{floweos1,floweos2,floweos3,floweos4,floweos5}. 
High-precision heavy-ion experiments on pion and kaon 
dynamics~\cite{floweos2,kaos1} revealed more details of the in-medium 
hadronic properties~\cite{softeos4,softeos5,softeos6,softeos7,softeos8}. 
A soft nuclear EoS at high baryon densities up to 
$(2-3)$-times the saturation density was predicted from transport theoretical 
strangeness production 
analyses~\cite{softeos4,softeos5,softeos6,softeos7,softeos8,softeos1,softeos2,softeos3}.

One realized soon a non-trivial density dependence of the EoS. The rise and 
fall of collective flow with increasing energy was a hint for sudden 
changes in the softness of the nuclear EoS at high 
densities~\cite{higheos1,higheos2,higheos3}. More than one decade passed 
to obtain more constraints on the EoS of highly compressed matter. 
The recent astrophysical measurements of neutron stars with masses of 
1.97 $\pm$ 0.04 $M_\odot$~\cite{NS1} and 2.01 $\pm$ 0.04 $M_\odot$~\cite{NS2} 
brought more controversial insights on the high-density EoS 
(see also the recent work by Fonseca et al.~\cite{fonseca} and the 
review article be Oertel et al.~\cite{oertel16}). 
These observations 
provide lower bounds for the maximum neutron star mass, excluding a soft 
EoS at high baryon densities. Neutron stars exhibit a complex internal 
structure~\cite{NS3,NS4a,NS4,NS5,NS6}. 
For instance, strangeness carrying mesons (kaons) and baryons (hyperons) 
can appear in the neutron star interior. 
In fact, the presence of hyperons ($\Lambda, \Sigma, \Xi$- and $\Omega$) 
inside neutron stars is in principle energetically allowed, since their 
chemical potentials are sufficiently large at high baryon densities. 
Nuclear matter with hyperons weakens the EoS largely at 
high baryon densities. Several nuclear models, successfully applied to nuclear 
systems (finite nuclei, heavy-ion collisions), cannot explain 
the observed data of neutron star masses, if they include hyperons in their 
descriptions. This is known as the "hyperon puzzle" issue. A recent and nice 
review on this controversial topic can be found in Refs.~\refcite{eoshyp1,eoshyp2}. 

It is thus important to understand better the strangeness part of the hadronic EoS. 
This is a difficult task due to the lack on detailed experimental information. 
In fact, for the nucleon-nucleon (NN) interaction (strangeness content $S=0$)
high precision $4300$ scattering data 
are available. They allow an accurate determination of the in-medium 
NN-interaction, at least for densities close to saturation. Adding 
strangeness to the system ($|S|=1$) the situation changes largely. 
In the $S=-1$-sector (hyperon-nucleon) one has so far had access to $38$ scattering 
data only. They still allow a reasonable determination of the $S=-1$ 
hyperon-nucleon (YN) interactions, however, with remaining ambiguities particularly 
for the in-medium YN-potentials. The higher strangeness domain, $S=-2$, is still 
an unobserved experimental region. Only theoretical predictions are available in 
the literature. Finally, the $S=-3$ sector concerning $\Omega N$-interactions has 
been up to present rarely explored. 

The YN-interactions are formulated on a group-theoretical basis, such as 
SU(3) symmetry. Besides SU(3) often SU(6) symmetry is employed. 
There exist phenomenological Skyrme-like approaches~\cite{skyrme} and 
covariant models. The latter are based on the Relativistic Mean-Field (RMF) 
theory, firstly introduced by D\"{u}rr~\cite{duerr} and Walecka~\cite{qhd1}, 
considerably improved by Bodmer and Boguta~\cite{qhd2} and further developed by 
Serot and Walecka~\cite{qhd3}. This RMF framework has been extended to the 
description of nuclear matter with hyperons~\cite{rmfhyp1,rmfhyp2,rmfhyp3}. 
Alternatively one can use RMF with density dependent coupling constants to mimic 
the microscopic non-linear structure of the interaction. The Density 
Dependent Hadronic (DDH) approaches~\cite{ddh1,ddh2,ddh3} have been applied to matter 
with hyperons too~\cite{ddhyp1,ddhyp2}. The microscopic approaches 
consider higher order correlations in the spirit of the ladder approximation. They are 
known in the literature as non-relativistic Brueckner-Hartree-Fock~\cite{bhf1,bhf2} 
or covariant Dirac-Brueckner-Hartree-Fock models~\cite{dbhf}. They are based on the 
One-Boson-Exchange (OBE) picture of the baryon-baryon interaction~\cite{obe1,obe2}. 
Further theoretical works toward a description of the full octet and decuplet 
baryons exist. Prominent examples for the $|S|=1$ YN-interaction 
are the models of the Nijmengen~\cite{YNa1,YNa2,YNa3,YNa4}, 
J\"{u}lich~\cite{YNb1,YNb2,YNb3} and Kyoto-Niigata~\cite{YNc1,YNc2,YNc3} groups. 
Recently, QCD inspired approaches have been also developed and lattice QCD 
simulations have been applied to the YN-interactions~\cite{YNd1,YNd2,YNd3}. 
Chiral effective field (EFT) theory has been also used for the construction 
of realistic in-medium YN-potentials~\cite{YNe1,YNe2,YNe3,YNe4}. 

The theoretical models for the YN-interactions, in free space and inside the hadronic 
medium, provide the essential physical input for transport studies of in-medium 
reactions induced by heavy-ions or hadrons impinging on nuclear targets. In relativistic 
collisions between heavy-ions or hadrons and nuclei one can probe the in-medium 
dynamics of produced hyperons and hyperfragments. Thus, such reaction studies 
can give useful information on the in-medium YN interaction by a systematic 
comparison with experimental data of produced hyperons and bound hypernuclei. 
Hypermatter is here of particular importance. While in 
single-$\Lambda$ hypernucleus the $|S|=1$ sector of the YN-interaction can be 
investigated, hypernuclei with higher strangeness content can be used to explore 
the higher strangeness part of the in-medium hyperonic interactions. Nuclei 
with more than one bound hyperons are referred to as multi-strangeness 
hypernuclei or superstrange nuclei. The denotation "superstrange" and 
the idea of looking at superstrange 
nuclei goes back to the pioneer work of Kerman and Weiss~\cite{hypfirst} with 
first estimations on multi-$\Lambda$ hyperfragment yields.

Hadronic in-medium collisions are described theoretically by kinetic theory, 
introduced by Ludwig Boltzmann in the year 
1872~\cite{phd51}. Boltzmann formulated a transport equation for classical 
many-body systems based on Liouville's theorem. Vlasov extended the Boltzmann 
equation by including a mean-field potential, followed by the famous work 
of Uehling and Uhlenbeck taking Pauli blocking effects in binary collisions 
into account~\cite{phd52}. Since then different theoretical aspects of transport 
dynamics were investigated. Kadanoff and Baym derived the kinetic equations from 
non-relativistic quantum statistics~\cite{baym}. Danielewicz introduced 
another microscopic derivation of the kinetic equations and applied them 
for the first time to the dynamics of heavy-ion collisions~\cite{dani}. 
Bertsch and Das Gupta continued the transport theoretical studies with 
more details concerning mean-field and collision dynamics~\cite{bertsch}. 
A modern covariant derivation of relativistic kinetic equations was 
formulated by Botermans and Malfliet~\cite{horror}, based on the microscopic 
Dirac-Brueckner-Hartree-Fock formalism. A more pedagogical introduction to 
relativistic kinetic theory can be found in De Groot~\cite{groot}. Since then 
various models based on (relativistic) kinetic theory have been applied in 
in-medium hadronic reactions. They are known in the literature as 
Boltzmann- Uehling-Uhlenbeck (BUU)~\cite{buu}, relativistic BUU 
(RBUU)~\cite{rbuu1,rbuu2,rbuu3}, Giessen-BUU (GiBUU)~\cite{gibuu}, 
Hadron-String-Dynamics (HSD)~\cite{hsd1,hsd2}, 
Quantum-Molecular-Dynamics (QMD)~\cite{qmd1,qmd2,qmd3}, relativistic 
QMD~\cite{rqmd1,rqmd2,bass}, Fermionic-Molecular-Dynamics (FMD)~\cite{fmd} 
and Antisymmetrized-Molecular-Dynamics (AMD)~\cite{amd}. A hydrodynamical 
description of heavy-ion collisions is also possible (see for details 
Ref.~\refcite{hydro}).

Nowadays transport theory is further developing to explore (among 
other tasks) in-medium single- and multi-strangeness 
dynamics~\cite{th1,th2,th3,th4,th5,th6,th7,th8,th9,th10,th11,th12,th13}. 
These studies have been motivated by a series of ongoing and forthcoming 
experimental activities, nicely reviewed in Ref.~\refcite{hyprev}. 
The Hyperon-Heavy-Ion (HypHI) collaboration~\cite{hyphi1} has recently 
reported longitudinal 
momentum spectra of low-mass single-$\Lambda$ hypernuclei in intermediate 
energy collisions between low-mass nuclei~\cite{hyphi2}. Within the J-PARC 
experimental project~\cite{jparc} high energy proton beams will be used for 
the production of bound hypersystems. We emphasize further experimental 
activities concerning hypernuclear studies, such 
as STAR (RHIC)~\cite{star}, ALICE (LHC)~\cite{alice}, FOPI and HADES at 
CBM~\cite{fopihades1,fopihades2} and NICA~\cite{nica}. 
The forthcoming \panda experiment at FAIR~\cite{panda1} is of great 
interest concerning multi-strangeness hypernuclear physics. Indeed, in-medium 
collisions with antiproton-beams at intermediate energies of few \GeV~only 
can overcome the high production thresholds of hyperons. The 
high annihilation cross sections at low incident energies into multiple meson 
production (antikaons) and the formation of strangeness resonances can 
accumulate energy and strangeness content via secondary scattering. Thus, a 
copious production of heavy hyperons through a multi-step collision process is 
possible. 

We review in detail the theoretical description of in-medium 
production of hypermatter in hadron- and ion-induced reactions. 
Section 2 deals with the theoretical details of the 
superstrange transport dynamics including the treatment of fragment and 
hyperfragment formation. Results of transport theoretical calculations are 
given in Sections 3 and 4. In Section 3 we discuss the production of hypermatter 
in heavy-ion collisions, while Section 4 is devoted to the formation of 
multi-strangeness hypersystems in hadron-induced reactions. Section 5 
summarizes this review. 

\section{Transport theoretical description of hadronic reactions}

There are several ways to derive the kinetic equations. 
One is based on the field-theoretical covariant 
description of strongly interacting many-body 
systems~\cite{green1,green2}. 
It starts from the non-equilibrium Green's functions of a many-body 
system. The various many-particle Green's functions are connected with 
each other through Dyson equations. Restricting up to $2$-particle 
correlations, one can derive within the Schwinger-Keldysh formalism 
relativistic kinetic equations for a correlated Green's function, which 
is related to a single-particle phase-space density. Further kinetic 
equations are obtained for spectral Green's functions, ending up with 
four coupled kinetic equations. The advantage of this derivation is 
a direct connection to the microscopic Dirac-Brueckner theory~\cite{horror} 
and a natural interpretation of the approximations (semi-classical limit 
and quasi-particles). 

Another simpler derivation is based on Quantum Hadrodynamics, 
which we follow here~\cite{rbuu3}. The advantage of this method 
is the direct relation to the more practicable mean-field theory of 
QHD~\cite{qhd3}. The starting point is the QHD Lagrangian density 
\begin{align}
{\cal L}_{QHD} = & 
	\overline{\Psi}\gamma_{\mu}
\left( i\partial^{\mu}-g_{\omega}\omega^{\mu} - 
g_{\rho}\vec{\tau}\,\vec{\rho\,}^{\mu}\right) \Psi 
-  \overline{\Psi}(m-g_{\sigma}\sigma-g_{\delta}\vec{\tau}\,\vec{\delta})\Psi
\nonumber\\
+ & {\cal L}_{\sigma}+{\cal L}_{\omega}+{\cal L}_{\delta}+{\cal L}_{\rho}
\label{QHD}
\,.
\end{align}
It includes the free Lagrangians for the Dirac spinors, which is explicitly 
given, and the free Lagrangians of the exchange $\sigma-,\omega-,\rho-$ and 
$\delta-$mesons. The mesonic degrees of freedom differ from each other in 
their internal Lorentz structure. The $\sigma$-field is a Lorentz-scalar, 
iso-scalar meson, the $\omega$-field is a Lorentz-vector, iso-scalar meson, 
the $\rho$-field is a Lorentz-vector, iso-vector meson, and finally, the 
$\delta$-field is a Lorentz-scalar, iso-vector meson. 
These mesons are responsible for the interaction between the spinors 
in the spirit of the OBE-model~\cite{obe1,obe2}. 
In the following derivations the $\rho$- and $\delta$-contributions will be omitted 
for simplicity. The meson field operators are replaced with their 
classical expectation values and constitute the classical mean-field (or Hartree) 
potential between the quantal spinor fields. From Eq.~(\ref{QHD}) one obtains  
the entire information for a physical system in the mean-field approximation. 
The energy and pressure 
densities are defined through the conserved energy-momentum tensor, i.e. the EoS. 
Furthermore, the Euler-Lagrange equations for the different degrees of 
freedom are derived, the Klein-Gordon and Proca equations of motion for the 
virtual mesons and the Dirac equation for the spinors. Latter is used for 
the derivation of the transport equation and reads
\begin{align}
\gamma_{\mu}
\left( i\partial^{\mu}-g_{\omega}\omega^{\mu} \right) 
\Psi 
-  (m-g_{\sigma}\sigma)\Psi = 0
\label{Dirac}
\,.
\end{align}

The derivation of the transport equation starts from the Wigner function
\begin{equation}
F_{\alpha\beta} (x,k) = 
\frac{1}{(2 \pi)^4} \int d^4R e^{-i k_{\mu}R^{\mu}} 
< \hat{\overline{\Psi}}_{\beta}(x+\frac{1}{2}R) \hat{\Psi}_{\alpha}(x-\frac{1}{2}R) >
\label{wigner}
\quad .
\end{equation}
This is the so-called Wigner-transformation of the correlated Green's function, 
which will be related with the $1$-body phase-space distribution $f(x,k)$. Note 
the matrix structure of the Wigner function, indicated in Eq.~(\ref{wigner}) 
by the greek subscripts in the spinors $\Psi$ and $\overline{\Psi}$. This 
matrix notation will be omitted in the following.

With the help of the definition of the Wigner function, Eq.~(\ref{wigner}) and 
the Dirac equation~(\ref{Dirac}) the following expression can be verified 
($x_{1} = x+\frac{1}{2}R$; $x_{2} = x-\frac{1}{2}R$)

\begin{eqnarray}
& \bigg ( \gamma_{\mu}(\partial^{\mu}-2ik^{\mu})+2im \bigg ) F(x,k) = 
\frac{2}{(2 \pi)^4} \int d^4R e^{-i k_{\mu}R^{\mu}} & \nonumber\\
& < \hat{\Psi}(x_1) \bigg ( \gamma_{\nu} \partial^{\nu}_{x_2} +
im \bigg)\hat{\Psi}(x_2) > &
\label{motion1}
\quad .
\end{eqnarray}
The right-hand side of Eq.~(\ref{motion1}) can be re-written with the help of 
the Dirac equation as 
\begin{equation}
\frac{2i}{(2 \pi)^4} \int d^4R e^{-i k_{\mu}R^{\mu}}
< \hat{\overline{\Psi}}(x_1) \bigg ( g_{\sigma} \hat{\sigma}(x_2)-
g_{\omega} \gamma_{\mu} \hat{\omega}^{\mu} (x_2) \bigg ) \hat{\Psi}(x_2) >
\label{motion2}
\quad .
\end{equation}
In the mean-field approximation the mesonic fields are just ${\cal C}$-numbers. 
Therefore, they can be taken out of the expectation values. Using 
the identity
\begin{equation}
g(x_2) = g(x-\frac{1}{2}R) = e^{-\frac{1}{2} R_{\mu} \partial^{\mu}_{x}} g(x)
\label{taylorg}
\end{equation}
the expression Eq.~(\ref{motion2}) can be transformed into the 
expression
\begin{equation}
\frac{2i}{(2 \pi)^4} \int d^4R e^{-i k_{\mu}R^{\mu}} \bigg [ 
e^{-\frac{1}{2} R_{\mu} \partial^{\mu}_{x}} \big ( g_{\sigma} \sigma(x)
-g_{\omega} \gamma_{\mu} \omega^{\mu}(x) \big ) \bigg ] 
< \hat{\overline{\Psi}}(x_1) \hat{\Psi}(x_2) >
\label{motion3}
\quad .
\end{equation}
Furthermore, using the identity
\begin{equation}
\int d^4R e^{-i k_{\mu}R^{\mu}} e^{-\frac{1}{2} R_{\mu} \partial^{\mu}_{x}} = 
e^{-i\frac{1}{2} \partial^{k}_{\mu} \partial^{\mu}_{x}} \int d^4 R 
e^{-i k_{\mu}R^{\mu}}
\label{hilfsgleichung}
\end{equation}
the right-hand side of Eq.~(\ref{motion1}) can be written as
\begin{equation}
2i e^{-\frac{1}{2} \partial^{k}_{\mu} \partial^{\mu}_{x}} \bigg (
g_{\sigma} \sigma(x) - g_{\omega} \gamma_{\mu} \omega^{\mu} (x) \bigg )F(x,k)
\quad .
\end{equation}
Thus, for the Wigner function $F(x,k)$ the following equation of motion is obtained
\begin{equation}
\bigg ( \gamma_{\mu} (\hbar \partial^{\mu} - 2ik^{\mu}) + 2im \bigg )F(x,k) = 
2ie^{-\frac{1}{2} i \hbar \Delta} \bigg ( g_{\sigma} \sigma(x) - g_{\omega} \gamma_{\mu} \omega^{\mu} (x) \bigg ) F(x,k)
\label{motion4}
\,,
\end{equation}
with the triangle-operator $\Delta$ defined by 
$\Delta \equiv \partial^{k}_{\mu} \partial^{\mu}_{x}$. 

The $\hbar$-constant has been explicitly included in Eq.~(\ref{motion4}) to 
understand better the semi-classical prescription. This approximation 
is manifested in a Taylor expansion of the exponential function in 
Eq.~(\ref{motion4}) up to first order in $\hbar$, which requires smooth 
behavior of fields and Wigner function in phase space. This approach is 
semi-classical in the sense, that the mean-field dynamics is treated 
classically, while quantal effects are included in the collision integral 
via Pauli blocking factors. The classical treatment is justified for 
reaction energies close to the Fermi energy and above, where the de Broglie wave 
length is small compared to the considered scales of few \fm. 

A Taylor expansion up to first order of Eq.~(\ref{motion4}) leads to the 
following expressions
\begin{equation}
[ \gamma_{\mu} k^{\ast \mu} - m^{\ast} ] F(x,k) = 0
\label{imaginaerteil}
\end{equation}
and
\begin{equation}
\bigg ( \gamma_{\mu} \partial^{\mu} - \Delta (  \Sigma_{s}(x) - \gamma_{\mu}  \Sigma^{\mu}(x) ) \bigg ) F(x,k) = 0
\label{realteil}
\end{equation}
for the imaginary and real parts of Eq.~(\ref{motion4}), respectively. 
Here in-medium self-energies were introduced, which contain the mean-field potential. 
The Lorentz-vector ($\Sigma^{\mu}$) and the Lorentz-scalar ($\Sigma_{s}$) parts 
define effective masses and kinetic momenta according 
\begin{eqnarray}
& k^{\ast\mu} = k^{\mu} - g_{\omega} \omega^{\mu} = k^{\mu} -  \Sigma^{\mu} &
\label{impuls_eff}\\
& m^{\ast}    = M - g_{\sigma} \sigma = M -  \Sigma_{s}
\label{masse_eff}
\,.
\end{eqnarray}

The imaginary part, Eq.~(\ref{imaginaerteil}), includes already the 
quasi-particle approximation. That is, the in-medium on-shell constraint 
for quasiparticles, which are characterized by an effective mass 
$m^{\ast}$ and a kinetic momentum $k^{\ast\mu}$. The real part, 
Eq.~(\ref{realteil}), is used to derive the transport equation (without 
collisions) for the phase-space distribution function. The 
expression~(\ref{realteil}) is still 
a matrix equation in spinor space. With a decomposition of the Wigner matrix 
$F_{\alpha\beta}$ into the elements of the Clifford-Algebra
\begin{eqnarray}
F(x,k) & = & {\cal F}(x,k) \cdot {\small 1}\hspace{-0.3em}1 + 
{\cal V}_{\mu}(x,k) \gamma^{\mu} + {\cal P}(x,k)\gamma_{5} + 
{\cal A}(x,k)\gamma_{\mu} \gamma_{5} \nonumber\\
 & + & {\cal T}^{\mu\nu}(x,k) \sigma^{\mu\nu} 
\label{clifford_algebra}
\end{eqnarray}
we obtain the desired transport equation for the scalar part 
${\cal F}(x,k)$, which can be identified with the $1$-body 
phase-space distribution function $f(x,k^{\ast})$
\begin{equation}
\Bigg ( k^{\ast}_{\mu} \partial^{\mu}_{x} + ( k^{\ast}_{\nu} 
F^{\mu\nu} + m^{\ast}(\partial^{\mu} m^{\ast}) ) \partial_{\mu}^{k^{\ast}} 
\Bigg ) f(x,k^{\ast}) = 0
\label{vlasov}
\quad .
\end{equation}
Eq.~(\ref{vlasov}) is known in the literature as the Vlasov equation. 
That is, the transport equation without the inclusion of binary 
collisions.  $F^{\mu\nu}$ is the field-strength tensor 
$F^{\mu\nu} = \partial^{\mu}  \Sigma^{\nu} - \partial^{\nu}  \Sigma^{\mu}$ 
and $f(x,k^{\ast})$ the phase-space distribution function. 

The Vlasov equation (\ref{vlasov}) describes the dynamics of the phase-space 
under the influence of a mean-field in terms of in-medium scalar and vector 
self-energies. This phase-space distribution fulfils Liouville's theorem 
($\tau$ is the eigentime)
\begin{equation}
\frac{d}{d \tau} f(x(\tau), k^{\ast}(\tau) ) = 
\bigg ( \frac{\p{x^{\mu}}}{\p{\tau}} \p^{x}_{\mu} + 
\frac{\p{k^{\ast\mu}}}{\p{\tau}} \p^{k^{\ast}}_{\mu} \bigg )
f( x(\tau), k^{\ast}(\tau) ) = 0
\label{liouville}
\end{equation}
for the invariance of the phase-space distribution. This 
theorem does not hold any more if collisions are taken into account. 
In binary processes particles can scatter out of a phase-space volume 
element (loss term) or into it (gain term). Both terms contribute to the 
collision integral, which can be derived in a consistent way using 
the Green's function formalism~\cite{horror}. Another way 
of derivation is based on the molecular chaos {\it Ansatz}~\cite{groot}. It assumes 
local collisions in space. The number of binary collisions is proportional to 
the corresponding $1$-particle phase-space distributions. The Pauli principle 
is taken into account. Combining these assumptions with the Vlasov equation 
we arrive at the relativistic transport equation. It is known in the 
literature as the Relativistic Boltzmann-Uehling-Uhlenbeck (RBUU) equation 
and reads
\begin{align}
& \Bigg ( k^{\ast}_{\mu} \partial^{\mu}_{x} + ( k^{\ast}_{\nu} 
F^{\mu\nu} + m^{\ast}(\partial^{\mu} m^{\ast}) ) \partial_{\mu}^{k^{\ast}} 
\Bigg ) f(x,k^{\ast})
\nonumber\\
& =  \frac{1}{2} \int \frac{d^3 k_{2}^{\ast}}{E^{*}_{2}(2\pi)^3} 
\frac{d^3 k_{3}^{\ast}}{E^{*}_{3}(2\pi)^3}
 \frac{d^3 k_{4}^{\ast}}{E^{*}_{4}(2\pi)^3} \, 
 W(k^{\ast}\, k_2^{\ast}|k_3^{\ast} \, k_4^{\ast})   
\nonumber\\  
& \times \; \Big[ \: f(x,k_3^{\ast}) f(x,k_4^{\ast}) 
		      \left( 1-  f(x,k^{\ast}) \right) 
		      \left( 1- f(x,k_2^{\ast}) \right) 
\nonumber \\  
 & \; \; \; \; \; \:
   -f(x,k^{\ast}) f(x,k_2^{\ast}) 
		\left( 1-  f(x,k_3^{\ast}) \right) 
		\left( 1-  f(x,k_4^{\ast}) \right) \: \Big] 
\label{rbuu}
\,,
\end{align}
with $E^{*}_{j}=\sqrt{m^{\ast\, 2}+k_j^{\ast\, 2}}$ for $j=2,3,4$. 
The right-hand side gives the temporal changes of the phase-space 
distribution $f(x,k^{\ast})$ due to binary collisions with other 
particles, if the Pauli principle allows it. The physical quantity 
for a scattering process is given by the transition probability 
$W(k^{\ast}\, k_2^{\ast}|k_3^{\ast} \, k_4^{\ast})$, which is 
proportional to the scattering cross section and to a $\delta$-function. 
Latter ensures energy-momentum conservation for each binary process. Here 
the formulation was given in terms of kinetic momenta. This is done only 
for technical reasons and simplifies the numerical calculations 
considerably~\cite{rbuu1}. One can use also the phase-space distributions 
in terms of kanonical momenta. Latter choice is preferable, if the 
mean-field is explicitly momentum dependent~\cite{weber,gibuu}. 

Numerically the RBUU equation is solved within the test-particle 
formalism~\cite{wong}, where the continuous phase-space is discretized 
by so-called test-particles. Any form of these test particles is allowed. 
Point-like ones lead, however, to numerical fluctuations in the calculation 
of densities and, thus, of the mean-field potential. It is more convenient 
to use a Gaussian form, as proposed by various authors~\cite{greg,rbuu1}. The 
computational procedure is considerably simplified by the fact of Liouville's 
theorem. From Eqs.~(\ref{vlasov}) and~(\ref{liouville}) one obtains 
the equations of motion for a test particle $i$ ($k=1,2,3$ are the spatial 
coordinates)
\begin{align}
\frac{d}{dt}x_i^{k} = & \frac{u_{i}^{ k}}{u_{i0}}
\label{dummy}\\
\frac{d}{dt}u_i^{\mu} = & \frac{1}{m^{\ast}_{i}u_{i0}} 
                       \bigg ( u_{i\nu}F^{\mu\nu}+\partial^{\mu}m^{\ast}_i
                      -(\partial^{\nu}m^{\ast}_i)u_{i\nu}u_i^{\mu} \bigg )
\label{prop}
\,,
\end{align}
with the $4$-velocity of particle with label $i$ given by 
$u^{\mu}_{i} = (\, u_{i0}, \vec{u}_{i} \,)$. It is related to the 
kinetic momentum via the relation $k^{\ast\mu} = m^{\ast}u^{\mu}$ 
and fulfils the in-medium on-shell condition 
$k^{\ast\mu}k^{\ast}_{\mu}=m^{\ast\, 2}$ or equivalently 
$u^{\mu}u_{\mu}=1$. 

At each time step one calculates the densities and mean-field potentials 
for each test-particle needed for the solution of the equations of 
motion~(\ref{prop}). Then the collision integral (the right-hand side of 
the RBUU equation) is numerically treated by a Monte Carlo method. The 
exclusive elastic and inelastic cross 
sections are used to determine the final channels. This final state can 
consist not only of $2$ particles, but it can be a multi-particle final 
state (see below). Then the modulus of the final state momenta is 
extracted from energy and momentum conservation, and their direction 
is determined from the differential cross sections. The whole procedure 
takes place in the local center of mass frame of the two colliding 
partners. The original method is explained in detail in Ref.~\refcite{gibuu}. 

Depending on the reaction type and the beam energy, several exclusive 
channels must be taken into account in the numerical treatment of the 
collision integral. For the simulation of heavy-ion collisions and 
proton-induced reactions up to incident energies of $1-2$ \GeV~the 
inelastic channels up to the $\Delta(1232)$ resonance including 
secondary scattering with pion production and absorption are sufficient. 
The formation of strangeness at these intermediate energies is a 
rare process. However, it should be included if one intends to study 
strangeness production. A more precise treatment of the 
collision term is possible by considering all baryonic resonances 
up to $2$ \GeV, which are given by the Particle Data 
Group~\cite{pdg}. This is realized in recent developments of 
transport, such as the UrQMD~\cite{bass}, HSD~\cite{hsd1} 
and the Giessen-BUU (GiBUU)~\cite{gibuu}. Latter realization we mostly 
used for the results of this article. More details 
on the cross sections of exclusive primary channels can be 
found in Ref.~\refcite{gibuu}. With increasing number of primary scattering 
processes, the number of 
secondary scattering increases too. Re-scattering is a very important 
mechanism for the formation of hypernuclei, as we will see 
later on. 

For the mean-field of nucleons different parametrizations exist in 
the literature. Within a non-relativistic transport equation 
Skyrme-type interactions are employed~\cite{skyrme}. Using the 
RBUU equation a covariant formulation of the nucleonic mean-field is 
obviously required. This is achieved by the various parametrizations 
in the mean-field theory of QHD, which exist in the literature 
too, see for instance~\cite{gibuu}. Essential here are those models with a rather soft 
EoS, which can describe the collective flow dynamics and kaon 
production fairly well~\cite{rbuu3,gibuu}. The mean-field of produced hyperons 
is constructed from the nucleonic one by SU(3) or SU(6) symmetry arguments. 

The mean-field for antiparticles within RMF should be treated with care. 
As discussed in Ref.\refcite{NLDbar} within a Non-Linear Derivative 
model~\cite{NLD}, the mean-field approach does not reproduce 
well the empirical energy dependence of the in-medium proton optical potential, 
particularly, at high energies. This issue becomes serious in the antiproton 
case. Indeed, standard RMF models fail to reproduce the empirically known 
regions of the in-medium $\bar{p}$-optical potential and diverge to infinity 
with increasing energy. Only a phenomenological re-scaling of coupling 
constants improves the situation~\cite{mishustin1,mishustin2}. As a standard 
procedure, transport models use re-scaled couplings for the antiparticle 
coupling constants. 

The transport equation gives the full information on single-particle 
dynamics of nucleons and produced particles. From the knowledge 
of the phase-space distribution one can determine thermodynamical 
properties such as particle densities, energy densities, pressure and 
the temperature respectively the excitation energy at 
each phase-space point of the dynamical evolution. 
\begin{figure}[th]
\begin{center}
\includegraphics[clip=true,width=0.6\columnwidth]{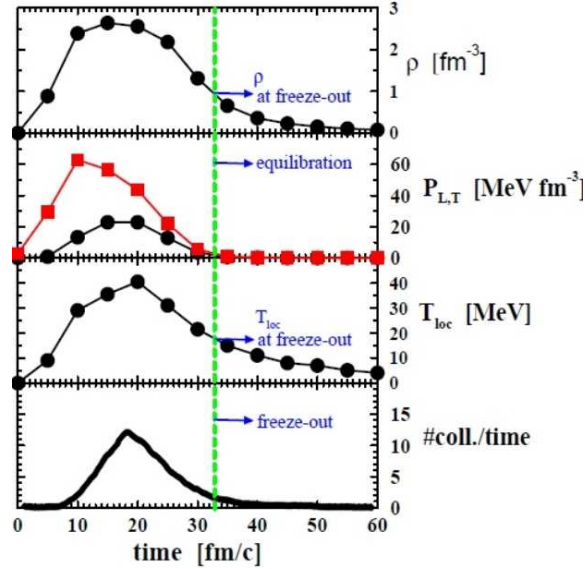}
\caption{\label{fig0}
Time Evolution of the density, longitudinal and transverse pressures 
(red and black curves in the second graph, respectively), temperature 
and number of collisions (from top to the bottom) at the central shell of a 
Au+Au heavy-ion collision at 0.6 \GeV beam energy per nucleon. The vertical 
line marks the onset of freeze-out. 
}
\end{center}
\end{figure}
This is 
very useful to gain information on (local) equilibration and 
the onset of instabilities. The degree of equilibration can be obtained 
by comparing the transverse and longitudinal components of the 
local energy-momentum tensor, called transverse and longitudinal pressure. 
The onset of an instability can be 
determined by calculating the pressure versus the density at a 
specific space-point as function of time. 
Fig.~\ref{fig0} shows the dynamics of thermodynamical quantities, extracted 
from RBUU calculations. They were obtained at the center of a heavy-ion 
reaction as function of time. The typical compression/expansion stages 
appear at intermediate times, as it can be seen from the time dependence of 
the central density and temperature. Both pressure components (longitudinal 
and transversal) differ from each other. That is, the participant system 
is out of local equilibrium. Freeze-out sets in when the particles do not 
collide any more. Just before freeze-out appears, 
the pressures are isotropic and local equilibrium occurs. 
In fact, as shown in 
Ref.~\refcite{essler}, spinodal instabilities appear at the center of 
participant and spectators at time stages close to freeze-out for 
heavy-ion collisions at intermediate incident energies. Thus, 
transport calculations can provide the onset of the fragmentation 
process, as the consequence of pressure instabilities. Physically 
this is explained as follows. The matter at densities below 
saturation intends to reach again the ground state, which is possible 
only by clusterization and formation of bound fragments. 

The fragmentation process can not be described by transport. The 
propagation of the physical fluctuations is missing, except the 
stochastic ones in the collision integral. There are attempts to go 
beyond the single-particle dynamics, see 
for instance Refs.~\cite{amd,colonna,ditoro}. Furthermore, the mean-field approach 
for the nuclear potential does not include clusterization processes 
for dilute matter. There exist recent developments towards 
this direction. Typel re-formulated the mean-field theory by considering 
in-medium clusters explicitly in the theoretical framework~\cite{typelnew}. 
In any case, with the knowledge of instabilities and equilibration from 
transport one can determine when fragmentation sets in and use, as 
an effective method, more sophisticated statistical models for the 
clusterization process. 

The Statistical Multi-fragmentation Model 
(SMM)~\cite{smm1,smm2} is a well-established approach to describe 
statistical de-excitation of residual systems. It includes the relevant 
mechanisms of fragment formation, i.e. evaporation, fission and multi-fragmentation 
as well as de-excitation of primary fragments. It has been used in 
hybrid simulations of in-medium hadronic reactions 
successfully, see for instance Refs.~\refcite{ogul1,ogul2,ogul3}. 
The connection between pre-equilibrium 
transport and statistical fragmentation models should be done 
with care. A critical quantity here is the excitation energy, 
which is obtained from the transport description of an excited 
source of hadronic matter, such as spectators in heavy-ion collisions 
or the residual nucleus in hadron-induced reactions. The excitation energy 
takes usually values of few \MeV~per nucleon only. That is, 
it is comparable with the binding energy per nucleon or even below 
it. On the other hand, the statistical models are strongly dependent 
on this quantity~\cite{smm1,smm2}, most likely due to the non-linear dependence 
on level densities. Thus, a very accurate determination of the 
excitation energy is required in transport studies for a reliable application 
of statistical fragmentation models. This task is closely related with a precise 
treatment of the initial stage in transport studies. The nuclei used 
for simulations should be initialized as precise as possible to 
avoid numerical noise in the temporal evolution of the binding 
energy and artificial particle emission. The binding energy enters 
into the calculation of the excitation of the residual source. The 
particle emission due to numerical fluctuations can affect 
the extracted mass and charge numbers of the source. 

This topic of an accurate initialization has been discussed in detail 
in Ref.~\cite{init}. Usually, empirical density profiles of ground state 
nuclei are used for the construction of initial configurations. This 
approach can be used for transport simulations at high energies, however, 
for the purpose of fragmentation with statistical models a better initialization 
procedure was developed. The density profiles are extracted from relativistic 
Thomas-Fermi (RTF)calculations by applying the same mean-field model as that used 
for the propagation of the initial configuration. Furthermore, the inclusion of 
surface effects in the binding energy and during the propagation is important 
to achieve a very good stability. 
\begin{figure}[th]
\begin{center}
\includegraphics[clip=true,width=0.6\columnwidth,angle=-90.]{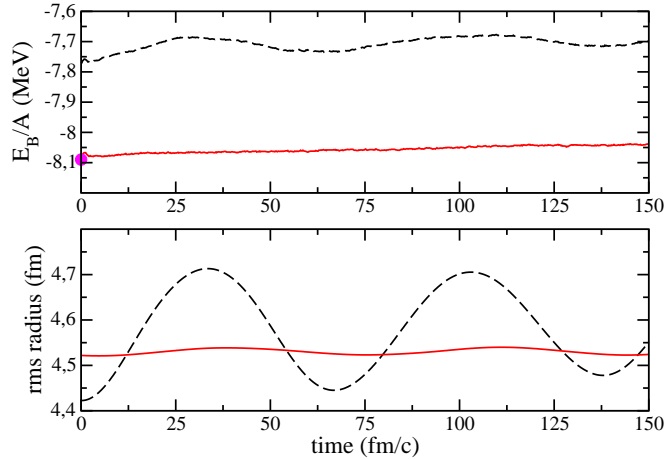}
\caption{\label{fig1}
Binding energy per nucleon (panel on the top) and rms radius (panel on the 
bottom) as function of time for a ${}^{112}$Sn nucleus. The solid (dashed) 
curves are obtained from Vlasov calculations using the improved (standard) 
initialization method. The filled circle symbol at $t=0$ \fm~marks the RTF 
value for the binding energy per nucleon.
}
\end{center}
\end{figure}
\begin{figure}[th]
\begin{center}
\includegraphics[clip=true,width=0.6\columnwidth,angle=-90.]{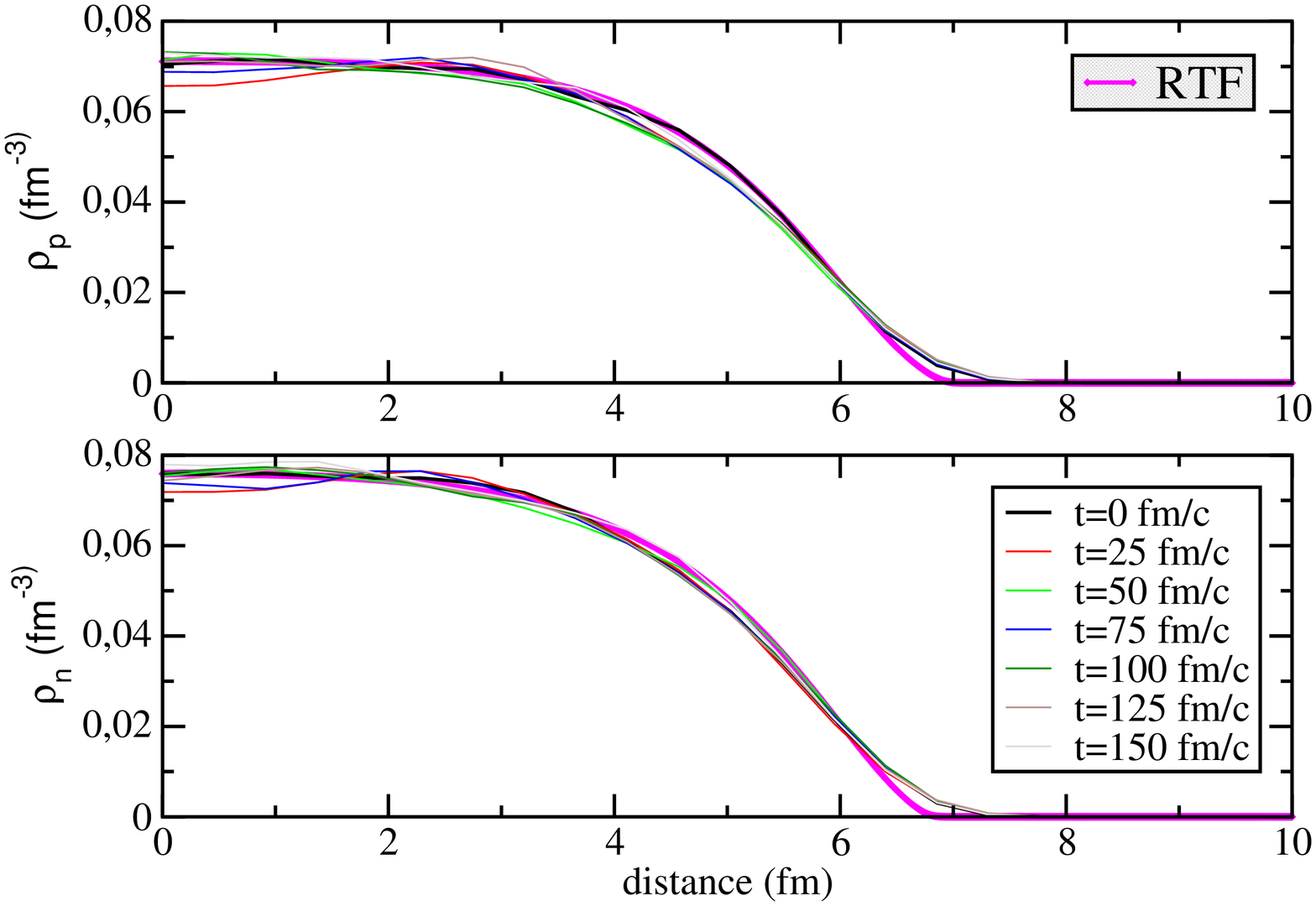}
\caption{\label{fig2}
Density profiles of protons (panel on the top) and neutrons 
(panel on the bottom) for a  ${}^{112}$Sn nucleus. 
The thick curve in both graphs indicates the RTF 
density distribution, while the other curves show the same quantity 
in the Vlasov calculations at different time steps, as displayed.
}
\end{center}
\end{figure}
As an example, we show in Fig.~\ref{fig1} the 
temporal evolution of the binding energy and root mean square (rms) radius 
of a single nucleus, as the result of a transport calculation using 1000 
test particles. The filled circle at $t=0$ \fm~is the RTF value. As one can 
see, the improved initialization prescription leads to an almost constant 
behavior of the binding energy, which is very close to the exact RTF value. 
Also the rms radius is stable in time, in contrast to the transport 
calculations with the conventional initialization method. A more detailed 
picture is given in Fig.~\ref{fig2} in terms of the density profiles. There 
the proton and neutron density distributions as function of the radial 
distance are shown. They have been extracted from Vlasov calculations at 
different times during the simulation of a single nucleus. The RTF reference 
densities of protons and neutrons are shown too. It is seen that the nucleus 
is very stable up to long time scales. Only at the surface the Vlasov 
densities are slightly wider relative to the RTF profile. This is due to 
the finite width of the Gaussian-shaped test particles. A more detailed 
discussion on this issue can be found in Ref.~\cite{init}. 

In the following sections we present and discuss the basic features of 
hypernuclei produced in collisions induced by heavy-ions and in 
antiproton-induced reactions. We have used the GiBUU model for these 
simulations. Other transport theoretical studies exist too. They will be 
discussed aiming to present the recent activities in this field of research.

\section{Strangeness dynamics in heavy-ion collisions}

\subsection{General features of reaction dynamics}

We first discuss the general method how pre-equilibrium transport 
and statistical approaches can be combined into a hybrid model. 
Proton-induced reactions are very well suited to test such hybrid models. 
At first, the target remains close to its ground state with moderate 
excitation. Furthermore, collective effects do not occur, in contrast 
to the violent compression and expansion dynamics in heavy-ion collisions. 
\begin{figure}[th]
\begin{center}
\includegraphics[clip=true,width=0.8\columnwidth]{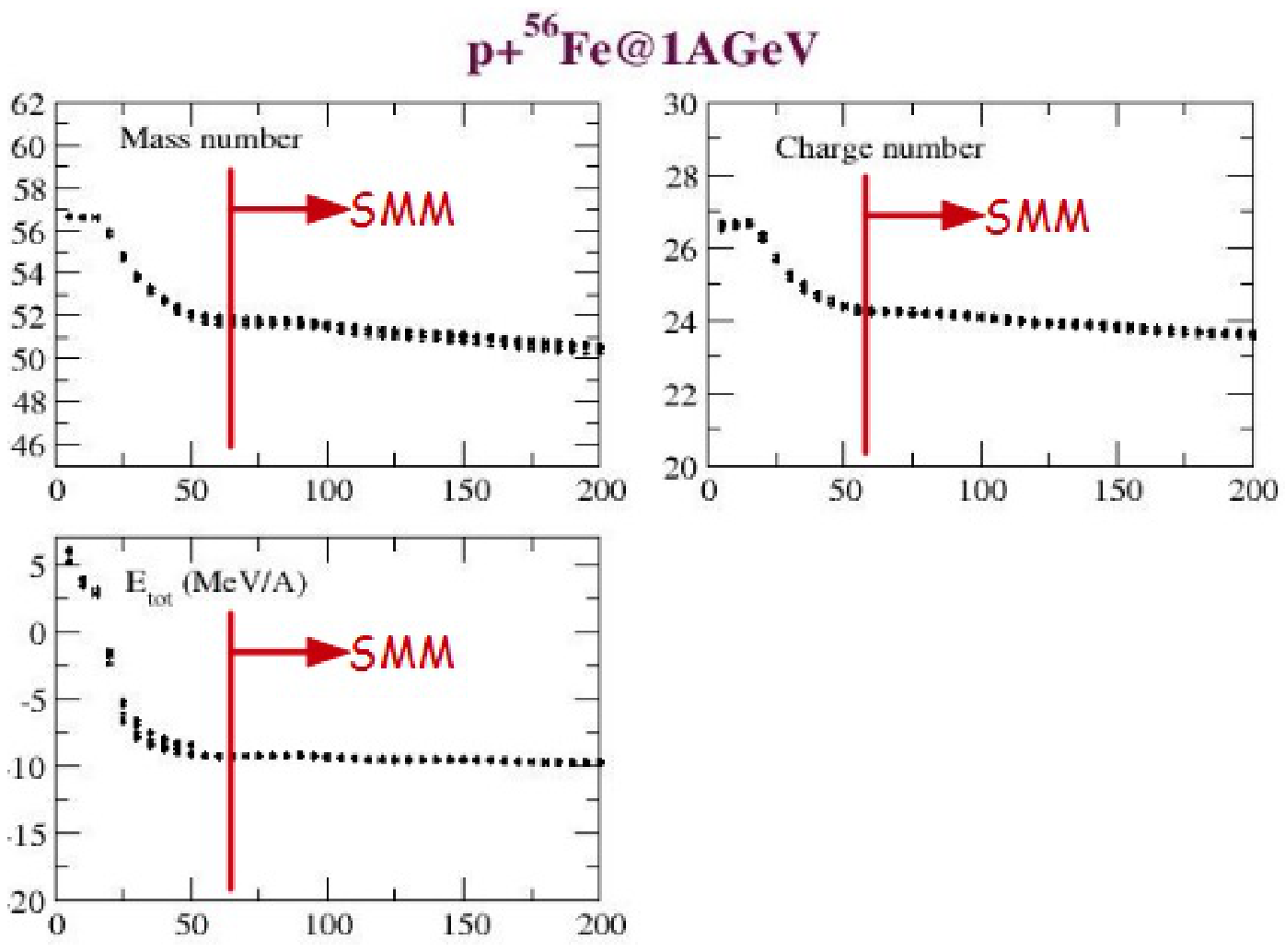}
\caption{\label{fig3}
Mass number, charge number and total energy per nucleon (less the nucleon rest mass) 
in the panels on the top-left, top-right and bottom-left, respectively, as function 
of time for the residual target nucleus. The vertical marks in each graph indicate 
the onset of the SMM model.
}
\end{center}
\end{figure}
According to the transport calculations, in a proton-nucleus reaction the 
nucleus gets excited by the proton beam. The nucleus starts to emit nucleons 
(pre-equilibrium emission) and a compound system is formed. 
As a residual source we define the compound system, which consists of all 
particles inside the nuclear radius by excluding the emitted nucleons. 
There are several methods to determine a residual source in transport simulations. 
One can use either the binding energy of the particles as criterion or apply 
a density constraint at each particle's position. Assuming that all nucleons inside 
the nuclear radius belong to the compound system, we define a residual (or fragmenting) 
source by the density constraint of $\rho_{cut}=0.01\times \rho_{sat}$. The particles 
with a density greater than $\rho_{cut}$ belong to the residual source. 
Fig.~\ref{fig3} shows the time evolution of the characteristic properties of the 
residual source. 
The high energy proton hits the 
target nucleus. During the beam penetration the target particles are excited 
due to subsequent collisions. At a time of $t_{fr}\simeq 60-70$ \fm~the residual 
target is de-excited due to pre-equilibrium particle emission, and the system 
approaches a freeze-out configuration. The residual system is still slightly 
excited, since its energy per nucleon (less the nucleon rest mass) is comparable 
with the ground-state binding energy per particle, but not the same. The difference 
between them gives the excitation of the system. Depending on the centrality of the 
proton beam, the value of the excitation energy takes values around $1$ \MeV~and 
below. The more peripheral the reaction, the smaller the excitation of the residual 
system. 

\begin{figure}[th]
\begin{center}
\includegraphics[clip=true,width=0.8\columnwidth]{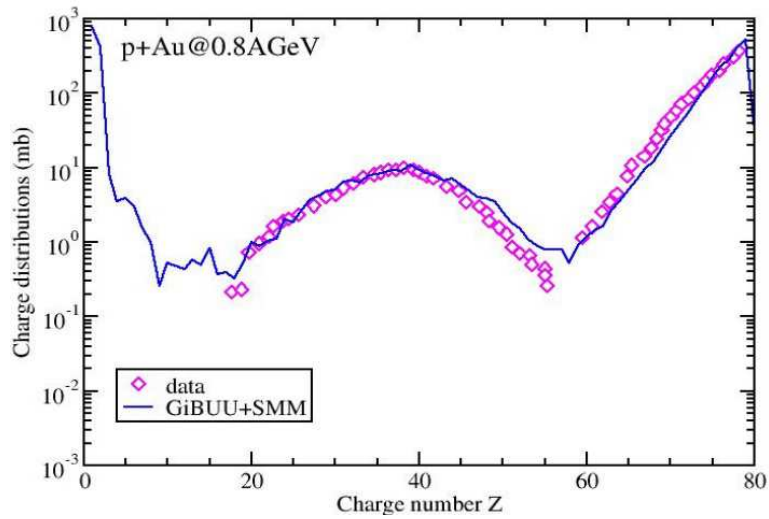}
\caption{\label{fig4}
Charge distribution for the reaction as indicated. Hybrid calculations 
(solid curve) are compared with experimental data (open symbols) taken 
from~\cite{data1a,data1b} (figure taken from~\cite{fig4ref}).
}
\end{center}
\end{figure}
\begin{figure}[th]
\begin{center}
\includegraphics[clip=true,width=1.0\columnwidth]{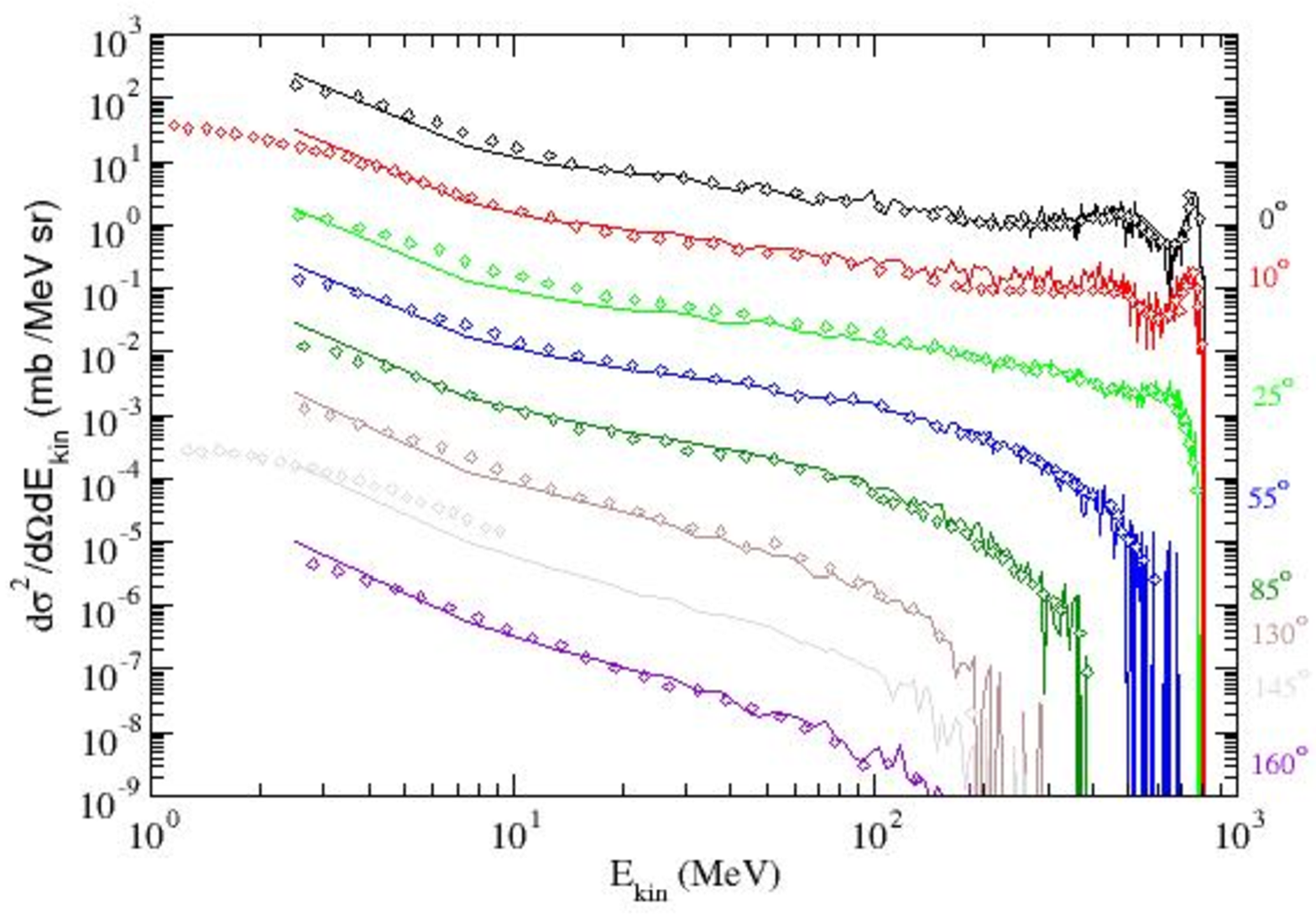}
\vspace{-5.0cm}
\caption{\label{fig5}
Kinetic energy spectra of emitted neutrons at different polar angles (as 
indicated on the right) for p+Pb reactions at 0.8 \GeV~beam energy. 
Hybrid calculations (solid curves) are compared with experimental data 
(open symbols) taken from~\cite{data2}. 
}
\end{center}
\end{figure}

We estimate the onset of the SMM model at the freeze-out time, 
i.e. at the time when a stable configuration has been reached. This is indicated 
in Fig.~\ref{fig3} by the vertical lines. Performing exclusive reactions 
for all the impact parameter 
range from central up to most peripheral collisions and applying at each simulation 
event the SMM model, one arrives to the results of Figs.~\ref{fig4} and~\ref{fig5}. 
There the charge distribution of fragments and differential energy spectra of free neutrons 
are shown. In particular, the charge distribution reproduces the experimental 
data fairly well. The evaporation peak close to the initial target charge number 
$Z_{init}=79$ and the wide fission peak at around $Z=79/2$ are clearly 
visible. These calculations predict also a multi-fragmentation region at low 
$Z$-values. This part of the distribution originates mostly from central events. 
For the formation of hypernuclei not only the mass and charge multiplicities, 
but also momentum distributions are crucial. The comparison of the hybrid 
calculations with data on differential energy spectra is shown in Fig.~\ref{fig5}, 
for a similar reaction system as in Fig.~\ref{fig4}. These spectra are selected at 
various polar angles and for free neutrons only. In this figure one realizes the 
importance of having both, pre-equilibrium and statistical emission. The high energy 
part of the spectrum with the clear visible quasi-elastic peaks at forward angles 
results from the particle emission of the transport calculations. The low energy 
spectrum of emitted particles is then the result of the SMM model due to de-excitation. 
The combination of both models, GiBUU and SMM, is required to explain the energy 
spectrum over the entire range.  

\begin{figure}[th]
\begin{center}
\includegraphics[clip=true,width=0.8\columnwidth]{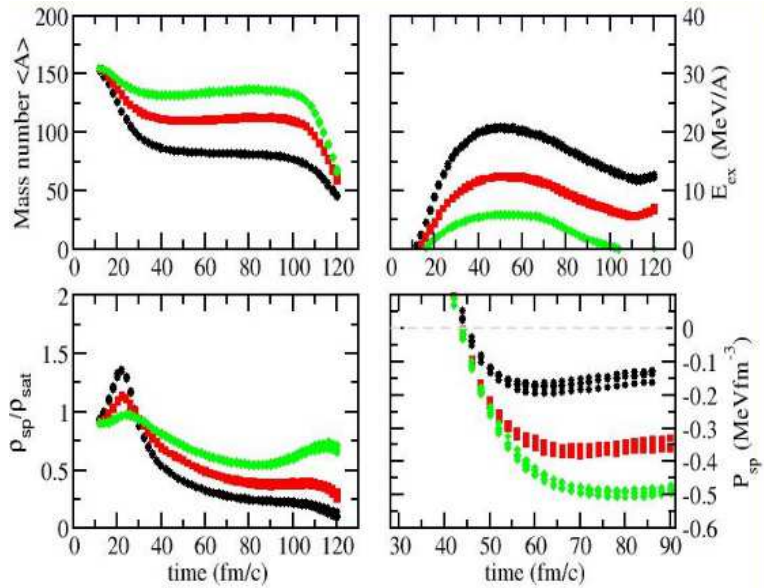}
\caption{\label{fig6}
Time evolution of the mass number (top-left), excitation energy per 
nucleon (top-right), central density relative to saturation (bottom-left) 
and central pressure (bottom-right) for projectile spectator in Au+Au 
collisions at 0.6 \GeV~beam energy per particle. The centrality increases 
from black to red and green for each panel. 
}
\end{center}
\end{figure}

The application of the combined approach in the dynamics of heavy-ion 
collisions follows in principle the same scheme. The difference with 
the case of proton-induced reactions are the 
collective effects. The participant region, which is formed during 
the pre-equilibrium stage, exhibits a violent compression/expansion 
dynamics showing up in a strong radial flow 
component~\cite{lisa,reisdorf}. Spectator dynamics, on the other hand, 
exhibits 
better controlled conditions. This is shown in Fig.~\ref{fig6}, where 
several properties of spectator matter are displayed as function of 
time. These are the mass number, the excitation energy, the density and 
the pressure (at the center of the projectile spectator). The three curves 
at each panel differ in the centrality. A similar situation appears as 
in the case of the residual nucleus in proton-induced reactions. After 
spectator's formation the system is firstly excited before cooling sets 
in. This can be seen in the graphs of Fig.~\ref{fig6}, where the mass 
number drops to a constant value. However, the system remains after cooling 
in an excited configuration. The degree of excitation is relatively high 
for semi-central collisions (black curves), while with increasing 
centrality (red and green curves) the excitation decreases. This is due to 
the less mixing between spectator and highly excited participant matter 
with rising impact parameter. Note that after ca. 50 \fm/c~the pressure 
becomes negative. This means the presence of fluctuations and the begin 
of the fragmentation process.

\begin{figure}[th]
\begin{center}
\includegraphics[clip=true,width=0.8\columnwidth]{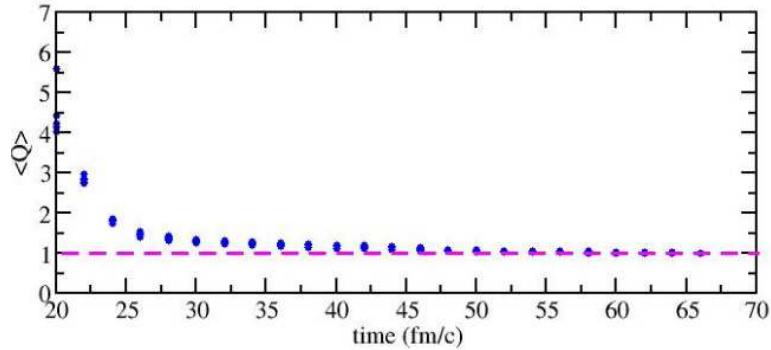}
\caption{\label{fig7}
Ratio between the two pressure components, longitudinal and transversal, 
for spectator matter for a semi-central Au+Au reaction at 0.6 
\GeV~incident energy per nucleon. 
}
\end{center}
\end{figure}

The question appears at which time step the SMM model should be applied. 
Here we refer to Fig.~\ref{fig7}, where the degree of equilibration is 
shown. In this figure the ratio between the longitudinal and transverse 
components of the central pressure in spectator matter is displayed as 
function of time. It is seen, that after roughly 50 \fm/c~the ratio approaches 
unity indicating the onset of local equilibration inside spectator 
matter. We have well defined conditions in spectator matter. That 
is, instabilities and equilibration occur at almost the same freeze-out 
time with a density of around $\rho_{sat}/3$ and a central 
temperature of $T\simeq 5$ \MeV~\cite{essler}. 

\begin{figure}[th]
\begin{center}
\includegraphics[clip=true,width=0.8\columnwidth]{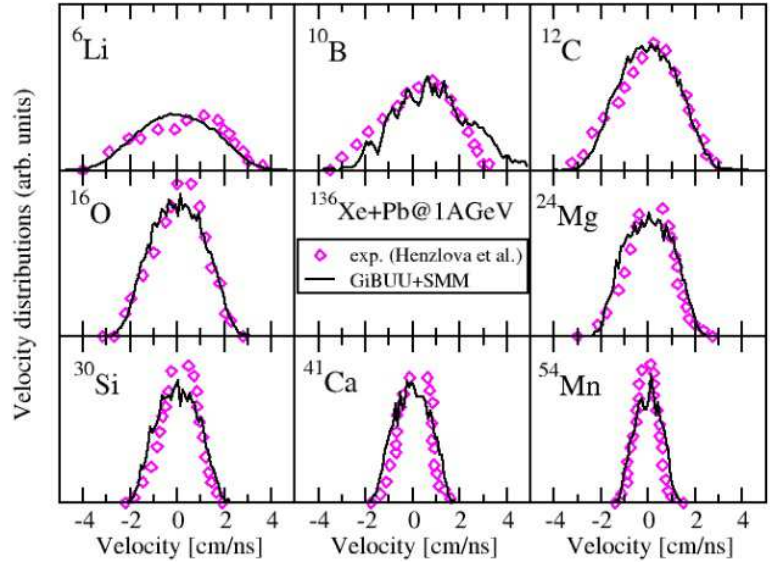}
\caption{\label{fig8}
Velocity distributions for various fragments at the rest-frame of 
projectile spectator, as indicated. The 
hybrid calculations (solid curves) are compared with experimental 
data (open symbols), taken from Ref.~\cite{data3} (figure taken 
from~\cite{th1}).
}
\end{center}
\end{figure}

Having the mass, charge numbers and the excitation energy of the spectator 
at freeze-out, one can apply the SMM model for spectator fragmentation. 
An example of this procedure is shown in Fig.~\ref{fig8}. There the 
velocity distributions of various spectator fragments for a 
Xe+Pb heavy-ion collision at 1 \GeV~beam energy per particle are shown. 
As one can see, the theoretical calculations reproduce the experimental 
data~\cite{data3} satisfactorily. More detailed comparisons including 
fragment multiplicities can be found in Ref.~\refcite{th1}. There exist 
also other recent studies on spectator fragmentation. In 
Refs.~\refcite{ogul1},~\refcite{ogul2} and~\refcite{ogul3} 
the SMM model has been applied to the fragmentation of projectile-like 
residues in intermediate energy collisions between Sn-isotopes. It is 
shown that the SMM approach reproduces the experimental isotope 
distributions fairly well. 

\subsection{Dynamics of strangeness and hypernuclei in heavy-ion collisions}

\begin{figure}[th]
\begin{center}
\includegraphics[clip=true,width=0.8\columnwidth]{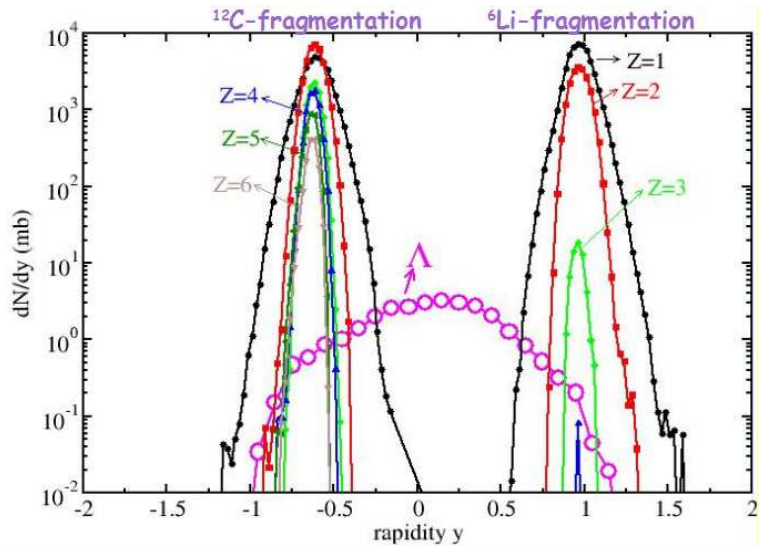}
\caption{\label{fig9}
Rapidity distributions of projectile and target fragments as well as 
of $\Lambda$-hyperons, as indicated, for Li+C heavy-ion collisions 
at 2 \GeV~incident energy per nucleon. 
}
\end{center}
\end{figure}

Two features should be described as precise as a model allows for 
a reliable production of hypermatter. That is, fragmentation and 
strangeness dynamics. Concerning fragmentation we have shown that this 
task can be well described by a combination of non-equilibrium transport 
and statistical approaches. Strangeness dynamics in heavy-ion collisions 
is consistently described with respect to data on kaons and 
hyperons~\cite{gibuu} (and references therein). We continue the discussion 
with the production 
of hypernuclei. Fig.~\ref{fig9} shows the rapidity spectra of produced 
fragments in projectile and target spectators as well as of hyperons. 
The fragment distributions are the result of the hybrid simulations, while 
the hyperons result from the transport calculations only. This figure shows 
the idea of coalescence for hypermatter production. In fact, a part of 
the hyperon spectrum overlaps with the fragment longitudinal momentum 
distributions close to projectile and target rapidities. 
The $\Lambda$-particles are mainly created in primary $BB\to BYK$-collisions 
and in secondary $\pi B\to YK$-scattering. Elastic and quasi-elastic 
scattering, i.e., with strangeness exchange, are included too. Secondary 
scattering involving fast pions is here important to create such a wide 
spectrum in beam momentum. Thus, some of the produced hyperons with a 
velocity close to that of spectators can be captured and form hyperfragments. 
This is realized by a coalescence in coordinate and momentum space between 
the hyperons and cold SMM fragments. 

\begin{figure}[th]
\begin{center}
\includegraphics[clip=true,width=0.9\columnwidth]{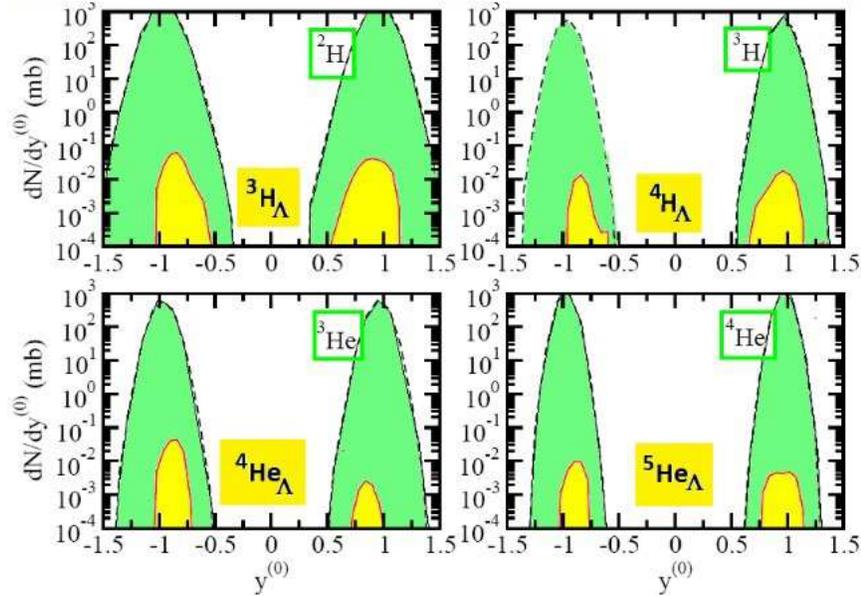}
\caption{\label{fig10}
Rapidity spectra of produced spectator fragments (dashed curves with green 
area) and the corresponding hyperfragments (solid curves with yellow 
area), as indicated. The considered reactions are a C+C collisions at 2 
\GeV~incident energy per nucleon.
}
\end{center}
\end{figure}
A typical result of hypermatter formation is shown in Fig.~\ref{fig10}. 
There the rapidity spectra of light-mass spectator fragments with 
the corresponding hyperfragments are displayed. The considered C+C 
reaction at 2 \GeV~beam energy per nucleon has been chosen, since for 
a similar colliding system (Li-beam on C-target) ongoing experimental 
activities are in progress~\cite{hyphi1,hyphi2}. 
On sees in Fig.~\ref{fig10}, that the light-mass hypernuclei are 
produced with relatively low cross sections. Their production rates 
are in the range of few $\mu b$ only. The main reason for the low 
production cross sections is the small size of the colliding systems. 
The smaller the nucleus, the less secondary scattering. Latter feature 
is, however, important to de-accelerate the hyperons inside the spectators, 
so that they can be captured. The predicted values of the light-hypernuclei 
are very close to the preliminary experimental data~\cite{hyphi2}. 
We have used here the symmetric C+C system for the determination of hypernuclei, 
in order to obtain a better statistics. 
In any case, it is desired to use colliding systems as heavy as 
experimentally possible, in order to obtain large hypernuclear 
cross sections. 

At present the study of hypermatter formation in intermediate energy 
heavy-ion collisions is an active field of research. There exist 
recent investigations by other groups. They use alternative approaches 
of transport dynamics combined with potential, coalescence and statistical 
prescriptions for the description of the fragmentation process. 
In Ref.~\refcite{th12} the well-established isospin-QMD transport model 
in combination with a newly developed fragmentation algorithm~\cite{th13} 
has been applied in collisions between heavy-ions. They have studied the 
formation of light-mass hypermatter in heavy-ion reactions at energies just 
above the kaon production threshold. As an important outcome of their 
analysis, it was found that rescattering strongly affects the hypernuclear formation. 
In Refs.~\refcite{th6} and~\refcite{th7} the Dubna-Cascade and 
the Ultra-relativistic QMD (UrQMD) kinetic approaches have been adopted for the 
dynamical treatment of heavy-ion collisions, in combination with potential and 
coalescence prescriptions for the formation of hypermatter. Their studies 
show the possibility of hypernuclear formation in heavy-ion reactions at 
intermediate energies. Even the production of multi-strangeness hypersystems 
is favorable with rather high cross sections, however, using heavy-mass 
projectile and target nuclei. For instance, single- and double-hypernuclei in 
spectator fragmentation are produced with cross sections of few \mb~and \mub, 
respectively, in Pb+Pb collisions around 1 \GeV~incident energy per 
particle. Bound hypermatter with $|S|=3$ is also predicted with 
cross section in the \mub-region at higher beam energies above 2 \GeV~per 
nucleon. The authors of Ref.~\refcite{th8} have used the UrQMD and 
HSD transport models in combination with a coalescence of baryons (CB) for 
the description of the in-medium hyperonic capture. In particular, 
they have applied the 
hybrid UrQMD+CB and HSD+CB approaches to collisions between heavy-ions of 
different size at projectile energies per particle above 2 \GeV~per 
nucleon. In their analysis the formation mechanism of hyperfragments 
originating not only from residual spectators, but also from the participant 
region, has been studied in detail. 

\section{Multi-strangeness dynamics in antiproton-induced reactions}

In in-medium $\bar{p}$-reactions strangeness particles are produced 
in annihilation processes close to the low-density surface region of 
the nucleus. Depending on the centrality, this perturbation can 
penetrate deep into the nucleus through multi-step binary processes. 
The average excitation per nucleon is comparable to that of the 
proton-nucleus reactions. 
\begin{figure}[th]
\begin{center}
\includegraphics[clip=true,width=0.8\columnwidth]{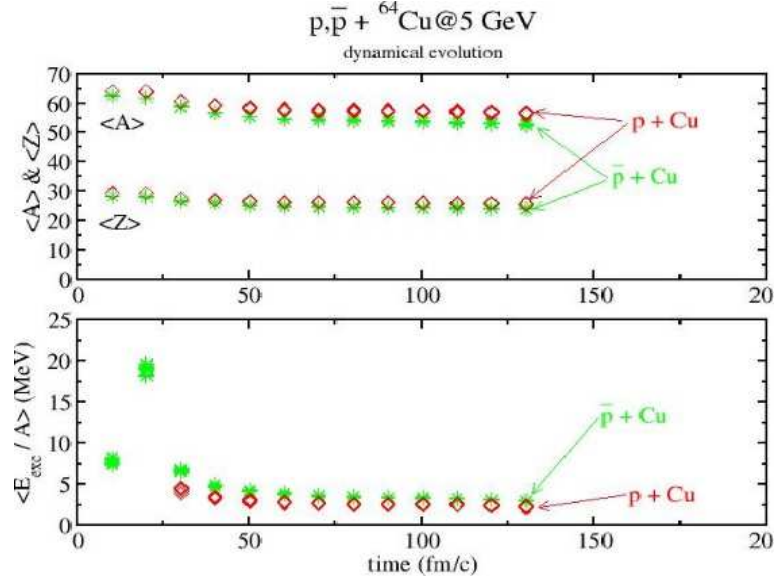}
\caption{\label{fig11}
Time evolution of the average mass number, charge number (upper
panel, as indicated) and the average excitation energy per nucleon (lower
panel) of the residual nuclei. GiBUU calculations for proton-induced (open diamonds)
and antiproton-induced (filled stars) reactions at an incident energy of 5 GeV and
impact parameter of b=3.4 fm are shown.
}
\end{center}
\end{figure}
This is shown in Fig.~\ref{fig11} in terms of the time evolution 
of average mass, charge and excitation energy per particle for both 
type of reactions. However, the main difference between $p$- and 
$\bar{p}$-induced reactions shows up in the abundance of 
produced particles, as it can be seen in Fig.~\ref{fig12}. Generally, 
the multiplicity of all produced particles increases in the 
$\bar{p}$-case. In particular, the multiplicity of antikaons and 
$\Lambda$-hyperons increases largely in the $\bar{p}$-induced reactions 
and the heavy $\Xi$-baryon appears. This is due to the strong 
annihilation cross sections mainly into multi-mesonic final 
states~\cite{larionov}. Furthermore, the annihilation into hyperon-antihyperon 
pairs becomes less and decreases by an order of magnitude with increasing 
mass of the produced (anti)hyperons. On the other hand, 
secondary scattering involving the cascade particles will be important for 
the formation of superstrange matter, as we will see.
\begin{figure}[th]
\begin{center}
\includegraphics[clip=true,width=0.8\columnwidth]{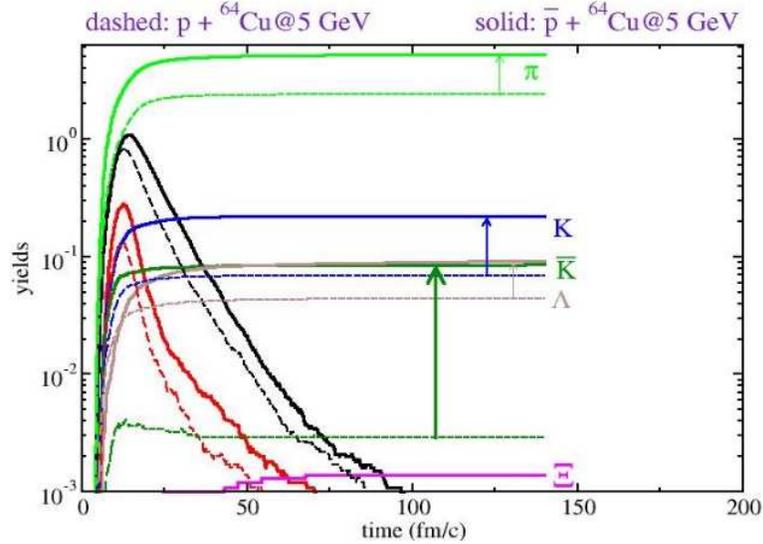}
\caption{\label{fig12}
GiBUU results for the particle yields as function of time for the same 
reactions as in Fig.~\ref{fig11}. The vertical arrows indicate the change 
of particle yields going from $p$-induced (dashed curves) to 
$\bar{p}$-induced reactions (solid curves). The different colored curves 
denote the various particles, as indicated. The black and red curves, which 
drop fast in time, correspond to $\Delta$ and higher resonances, respectively.
}
\end{center}
\end{figure}
The fragmentation process here (see Fig.~\ref{fig13}) is very similar to 
the $p$-induced reactions, as it can be seen in Fig.~\ref{fig4}. 
Again, evaporation, fission and 
multifragmentation regions are visible going from most peripheral to most 
central reactions. 
\begin{figure}[th]
\begin{center}
\includegraphics[clip=true,width=0.8\columnwidth]{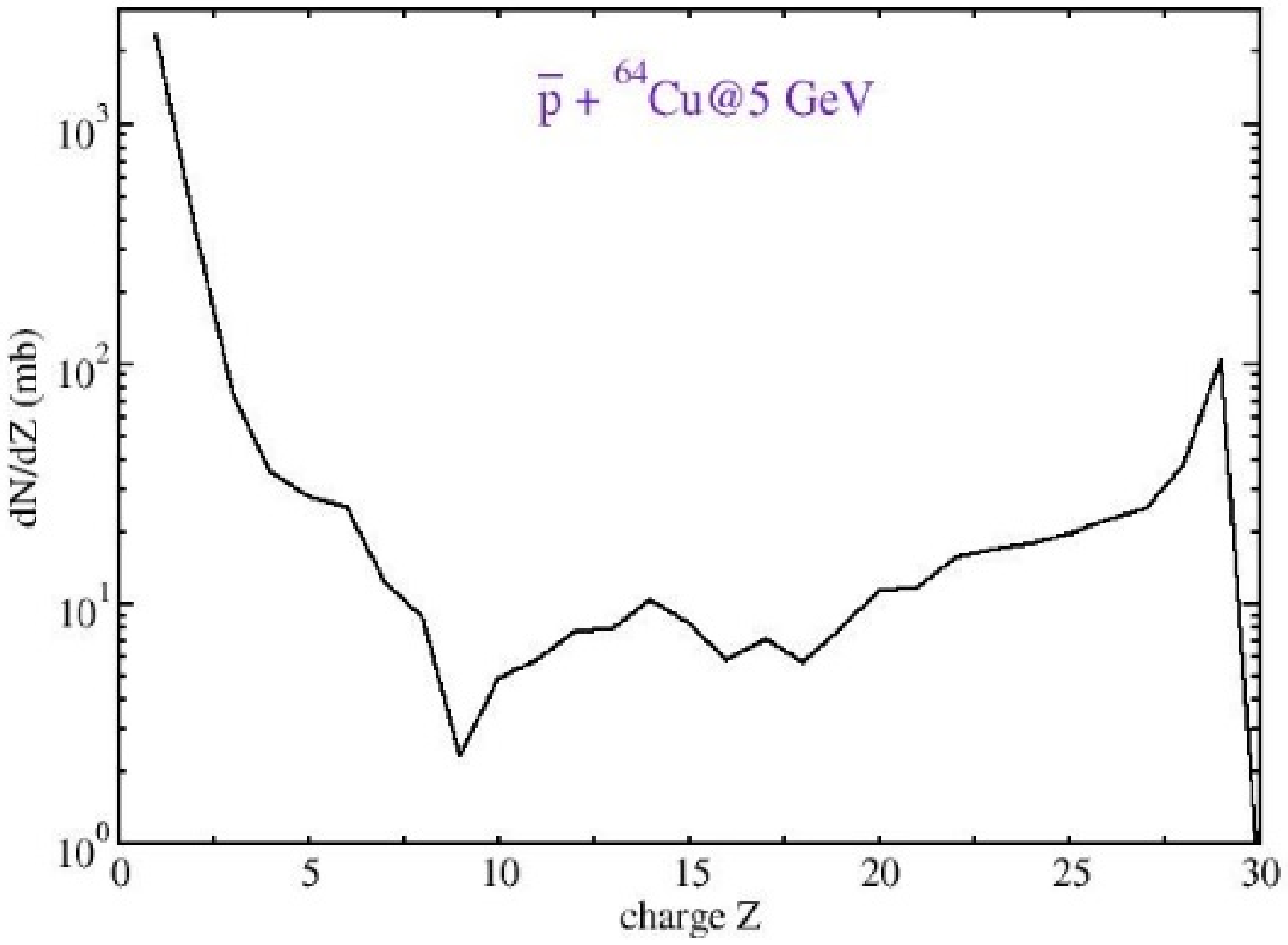}
\caption{\label{fig13}
Fragment charge distribution for centrality-inclusive $\bar{p}$-induced 
reactions, as indicated. The curve is the result of the 
GiBUU+SMM hybrid approach.
}
\end{center}
\end{figure}

\begin{figure}[th]
\begin{center}
\includegraphics[clip=true,width=0.8\columnwidth]{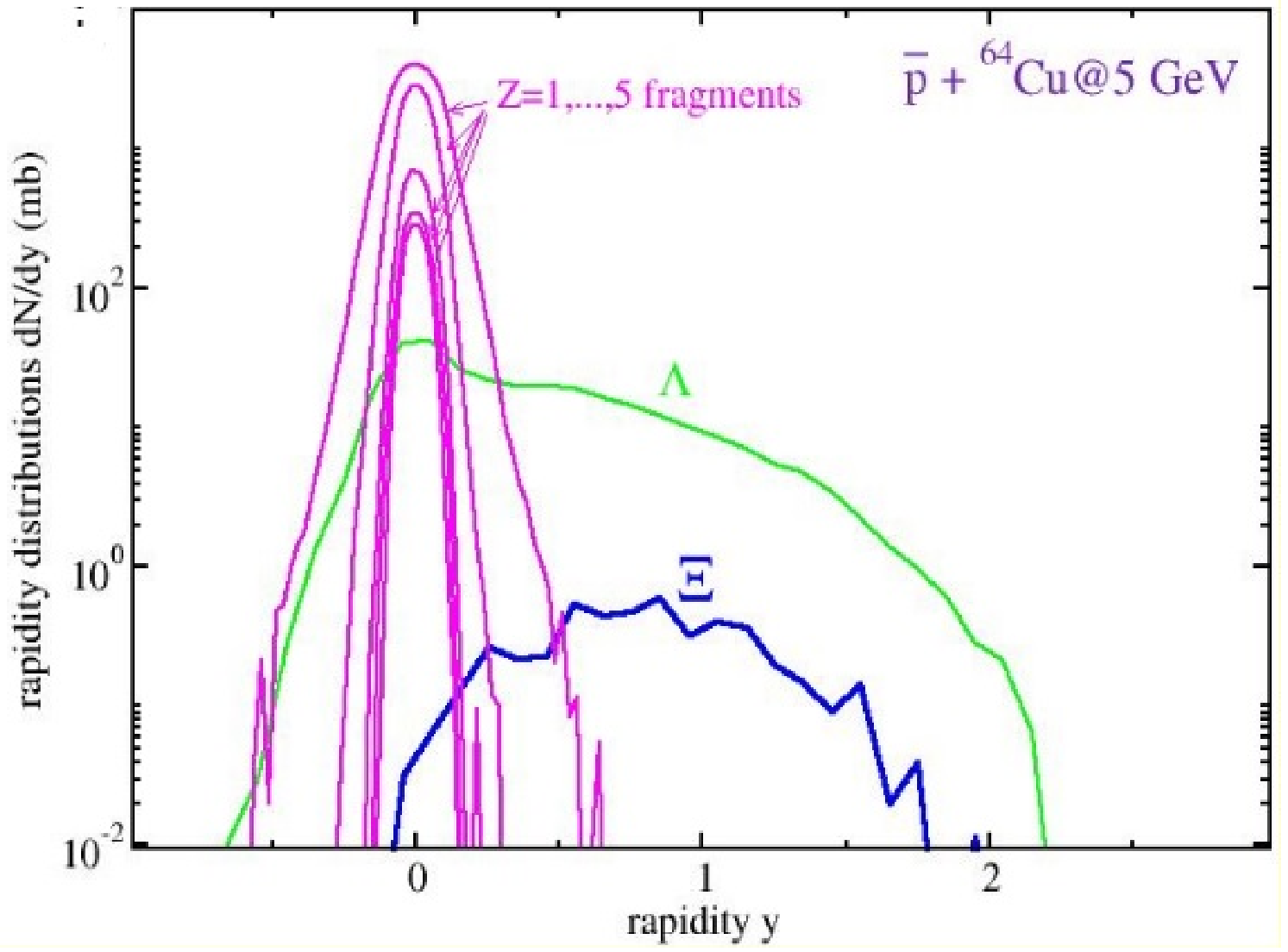}
\caption{\label{fig14}
Rapidity spectra of different fragments, $\Lambda$- and $\Xi$-hyperons for the 
same reaction as in Fig.~\ref{fig13}. The curves result from 
GiBUU+SMM hybrid calculations (figure taken from~\cite{th2}).
}
\end{center}
\end{figure}

Fig.~\ref{fig14} summarizes the results of the pre-equilibrium transport 
and SMM calculations in terms of the rapidity distributions of produced 
residual fragments and hyperons. The $\Lambda$-rapidity spectrum is rather 
broad. Secondary scattering is responsible for the low energy part of 
produced $\Lambda$-particles. This supports the formation of 
$\Lambda$-hypernuclei, as in the case of spectator fragmentation in 
heavy-ion collisions. The most interesting part here is the $\Xi$-production, 
which will be responsible for the production of multi-strangeness hypermatter. 
$\Xi$-particles are created with rather high probability, even if 
their production cross sections from annihilation are very low. In fact, 
while $\Lambda\bar{\Lambda}$-pairs are produced with cross sections of several 
hundred \mub, the antiproton annihilation cross section for $\Xi$-production 
is very low with orders of few \mub~only~\cite{kaidalov}. 

\begin{figure}[th]
\begin{center}
\includegraphics[clip=true,width=0.8\columnwidth]{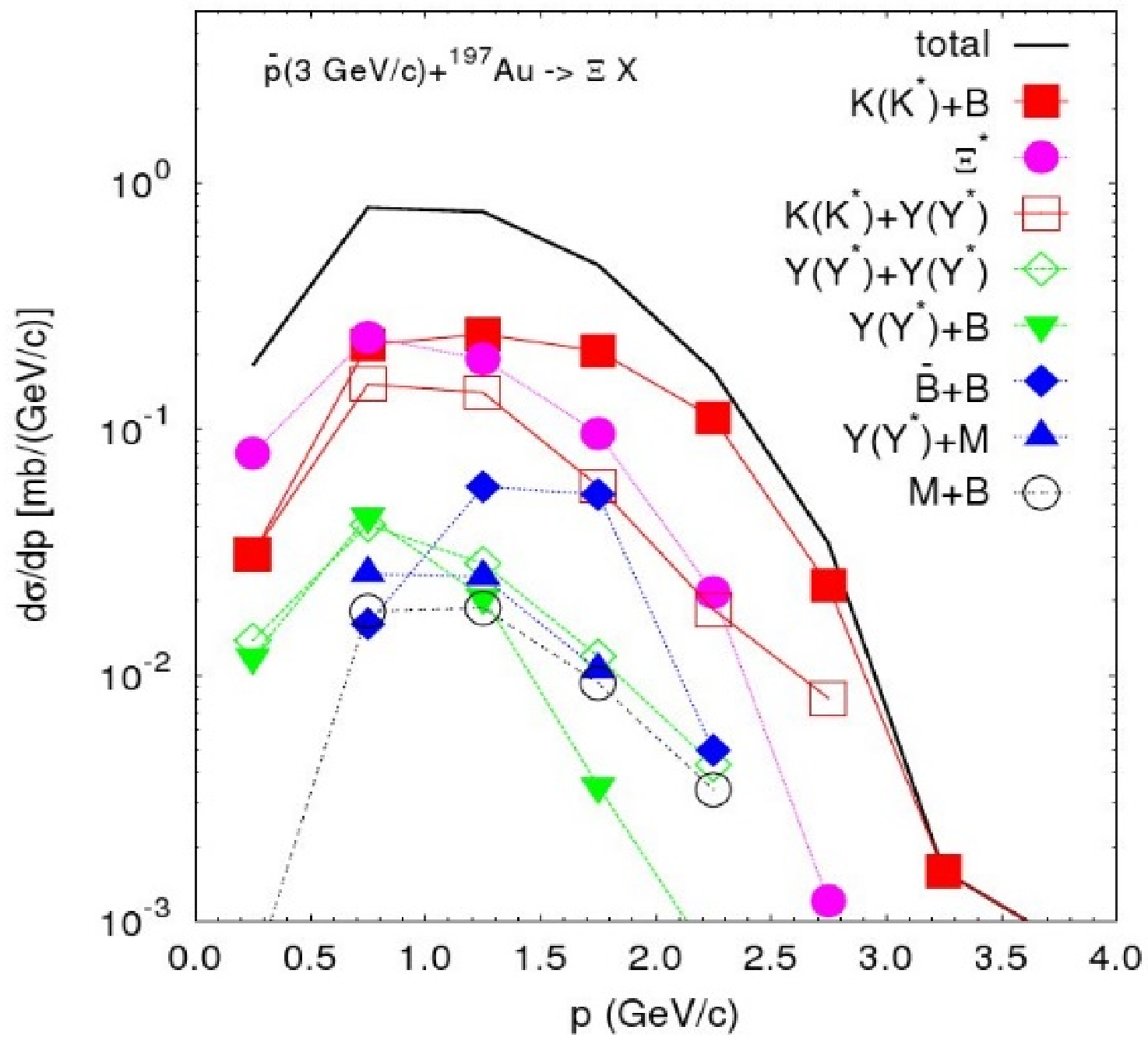}
\caption{\label{fig15}
GiBUU results for the momentum spectrum of inclusive $\Xi$-production (solid curve) 
together with the partial contributions to the total $\Xi$-production for the 
antiproton-induced reaction, as indicated (figure taken from~\cite{larionov}).
}
\end{center}
\end{figure}

Most of 
the $\Xi$-hyperons escape the nucleus, but there is a small fraction 
with rapidities close to those of the residual fragments.
Due to the high production threshold of the heavy $\Xi(1315)$-particles 
one would naively expect that they escape the nucleus with high rapidities. 
However, secondary scattering of produced $\Xi$-hyperons with the hadronic 
environment is crucial for low energy cascade particles. This is manifested 
in Fig.~\ref{fig15} in terms of momentum spectra of the total $\Xi$-yield 
including the exclusive contribution channels. Indeed, secondary scattering 
processes involving (anti)kaons, and kaonic/hyperonic resonances contribute 
largely to the low energy tail of the $\Xi$-momentum distribution. The situation 
is similar for the momentum spectra of produced $\Lambda$ and $\Sigma$ hyperons 
in $\bar{p}$-induced reactions, where experimental data exist~\cite{hyps1data}.

The formation of not only single-$\Lambda$ hypernuclei, but also of double-$\Lambda$ 
hypermatter is possible in $\bar{p}$-induced reactions. However, the probabilities 
of $|S|=2$-hypermatter are still very low with respect to $\Lambda$-hypernuclear 
yields~\cite{th2}. An alternative method has been proposed by the 
\panda-collaboration to enhance the production of superstrange bound 
matter~\cite{panda1}. That is a two-step reaction with primary and secondary targets. 
An antiproton beam interacts with a first target. The low energy part of the 
produced $\Xi$-hyperons can be used as a secondary beam for $\Xi$-induced reactions 
on the secondary target. The interaction between the $\Xi$-particles with the 
particles of the secondary target can create multiple captured hyperons and, thus, 
multi-strangeness hypernuclei. Indeed, transport-theoretical studies support this 
scenario~\cite{th2}. 

\begin{figure}[th]
\begin{center}
\includegraphics[clip=true,width=0.8\columnwidth]{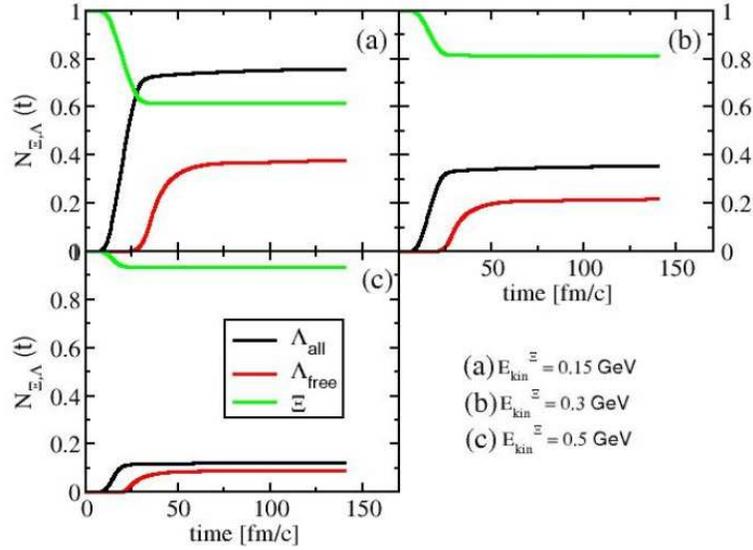}
\caption{\label{fig16}
GiBUU calculations for the total yields (normalized to unity at $t=0$ \fm/c) 
of $\Lambda$ (black curve) and $\Xi$ (green curves) particles as function 
of time for $\Xi$-induced reaction on Cu-target at three beam energies. The 
red curve show the yield of free $\Lambda$-hyperons only. 
}
\end{center}
\end{figure}

Fig.~\ref{fig16} shows the strangeness dynamics for $\Xi$-induced reactions at 
three different low energies of the $\Xi$-beam. Two aspects are visible here. 
At first, a significant capture of $\Lambda$-particles inside the matter is observed. 
Secondly, this feature shows a strong energy dependence. The main mechanism of 
$\Lambda$-production here is the inelastic $\Xi N\to \Lambda\Lambda$ channel. 
According to microscopic calculations~\cite{rijkenold,fujiold} this cross section can 
increase largely at low $\Xi$-energies with respect to elastic 
$\Xi N\to \Xi N$-scattering. Thus, double $\Lambda$ production drops 
with increasing energy of the cascade particles. This energy dependence in 
transport calculations 
is very strong and reflects just the strong energy dependence of the 
corresponding cross section~\cite{rijkenold,fujiold}. Note that for these important 
channels no experimental data exist, in contrast to 
$|S|=1$-scattering~\cite{YNb2,YNb3,YNc1,YNc2,YNc3}. 

\begin{figure}[th]
\begin{center}
\includegraphics[clip=true,width=0.8\columnwidth,angle=-90.]{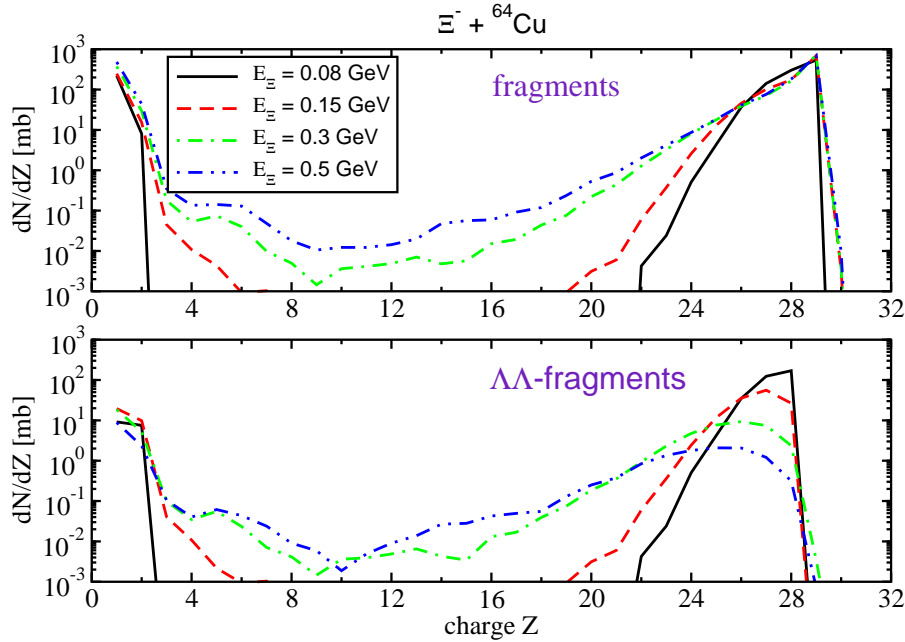}
\caption{\label{fig17}
Charge distributions of fragments (panel on the top) and double-$\Lambda$ 
hyperfragments (panel on the bottom) for $\Xi$-induced reactions at 
incident energies as indicated (figure taken from~\cite{th2}).
}
\end{center}
\end{figure}

Formation of double-strangeness $\Lambda\Lambda$ hypernuclei can thus occur 
in the secondary reaction. This is shown in Fig.~\ref{fig17}, where 
the charge distributions of fragments (upper panel) and 
$\Lambda\Lambda$-hyperfragments (lower panel) at different $\Xi$-energies are shown. 
At first, the fragment distribution becomes broader with increasing 
beam-energy of the cascade particles. Higher incident energy is associated 
with increasing excitation of the residual target nucleus. Thus, the 
fission region at around half the initial target charge and multi-fragmentation 
show up with rising beam energy. As an important result, an abundant production 
of $|S|=2$-hypernuclei is predicted by these transport calculations. 
This is visible in Fig.~\ref{fig17} by comparing corresponding curves for 
fragments (upper panel) and hyperfragments (lower panel) for each 
incicent $\Xi$-energy. With increasing beam energy the production of 
hypermatter around the evaporation peak drops significantly. However, 
the production yields of double-strangeness 
$\Lambda\Lambda$-hypernuclei are in the \mb-region at these low 
$\Xi$-energies.

Therefore, the proposed \panda-experiment can be very well suited to 
explore in more detail not only the nucleon-hyperon interactions, but 
also the still less understood regions of the higher strangeness sectors. 
Recent theoretical activities have investigated the in-medium 
hyperon interactions for $|S|>1$~\cite{YNa1,YNa2,YNa3,rijkenold,fujiold}. 
These are based on the microscopic meson-exchange picture or 
using quark-cluster approaches. The predictions between 
the theoretical models differ to a large extend. Therefore, 
a parameter-free theoretical framework would be obviously desired. This is 
possible only if one goes beyond the meson-exchange picture and considers  
the internal hadron structure. Recent attempts in this direction have 
been started within the more sophisticated Lattice 
QCD-simulations~\cite{YNd1,YNd2,YNd3} and chiral-EFT 
theories~\cite{YNe1,YNe2,YNe3,YNe4}.
\begin{figure}[th]
\begin{center}
\includegraphics[clip=true,width=0.8\columnwidth]{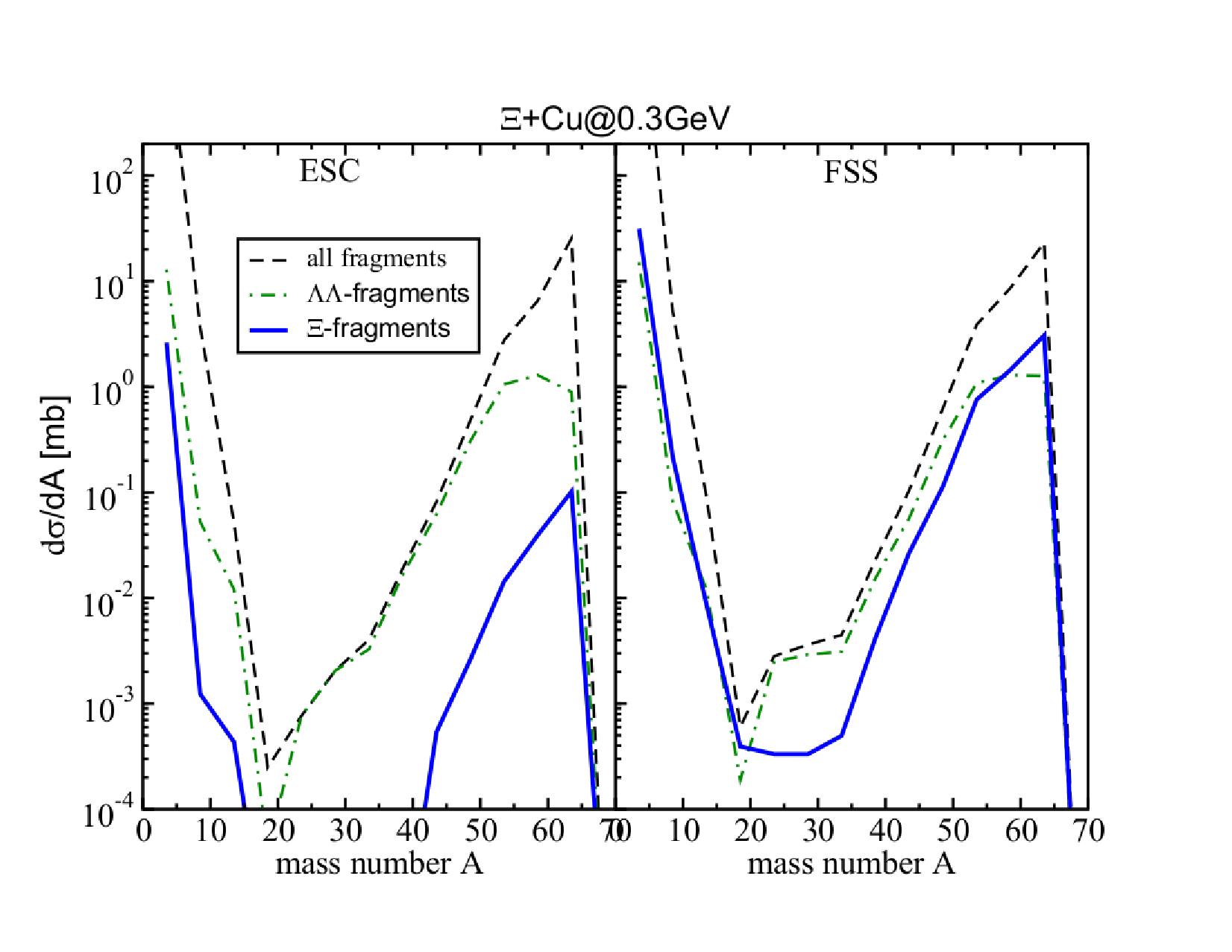}
\caption{\label{fig18}
Mass distributions of fragments (dashed curves), 
$\Lambda\Lambda$-hyperfragments (dot-dashed curves) and $\Xi$-hyperfragments 
(thick solid curves) for $\Xi$-induced reactions at 0.3 \GeV~incident energy. 
Both panels show the results of the GiBUU+SMM model using two different 
microscopic approaches for the $\Xi N$-interaction (figure taken from 
\cite{th3}).
}
\end{center}
\end{figure}
The formation of multi-strangeness bound matter in \panda-type reactions can 
be used as a probe to test the microscopic predictions for the superstrange 
sector of the in-medium hyperon potential. This issue has been studied 
recently in detail~\cite{th3}. An example is shown in Fig.~\ref{fig18}, where 
the mass number distributions of all fragments, double-$\Lambda\Lambda$ 
hyperfragments and, in particular, $\Xi$-hyperfragments are displayed using two 
different approaches for the $\Xi N$-scattering. The transport results on the 
left are based on the extended-soft-core (ESC) approach~\cite{rijkenold}, 
while those 
on the right are performed within the quark-cluster model (FSS)~\cite{fujiold}. 
Both models 
lead essentially to different results for $\Xi N$-elastic and inelastic 
$\Xi N\to\Lambda\Lambda$-scattering. In particular, the FSS 
$\Xi N\to\Lambda\Lambda$-cross section is strongly reduced relative to the 
ESC-predictions. Thus, the $\Xi$-multiplicity increases in the transport 
calculations using the FSS model. The consequence is a higher production 
yield of $\Xi$-hypernuclei, as clearly shown in Fig.~\ref{fig18}. Note that 
the formation of $\Xi$-hypernuclei beyond the conventional evaporation region 
is possible depending, however, on the microscopic model applied. 
The \panda-proposed scenario could thus be used to better constrain the 
higher strangeness sector of the in-medium interaction. 

\section{Summary and conclusions}

In summary, in-medium reactions induced by heavy-ions and antiproton-beams represent 
an excellent tool to study in more detail the multi-strangeness sector of the hadronic 
equation of state. The knowledge of the in-medium superstrange interactions is crucial 
not only for nuclear and hadron physics, but also for nuclear astrophysics. 
Furthermore, reaction studies on bound superstrange hypermatter offer great 
opportunities to explore the unobserved regions of exotic bound hypersystems. 

The transport-theoretical description of in-medium hadronic reactions is indispensable 
for hypernuclear studies. Microscopically developed approaches for in-medium interactions 
can be probed in complex situations of the reaction dynamics within kinetic approaches. 
Transport simulations can be also used to simulate the event structure of proposed 
experiments. A combination of pre-equilibrium dynamics and statistical fragmentation 
is a very useful tool to understand better the complete dynamics in such reactions, 
i.e., pre-equilibrium propagation and dynamical particle production as well as 
statistical fragmentation. 

In-medium hadronic reactions offer also other possibilities of study. By extending 
heavy-ion reactions to heavier colliding systems and to higher beam energies above 
the strangeness production thresholds, one can probe definitely superstrange matter 
at baryon densities far beyond saturation. Such a task is theoretically 
possible and experimentally feasible at the Compressed-Baryonic-Experiment (CBM) 
at FAIR. Furthermore, hadron-induced reactions, such as proposed by \panda, but using 
high energy secondary $\Xi$-beams can be useful to explore $|S|=3$-superstrange 
dynamics involving the heavy $\Omega(1673)$-baryon. The particular nature 
of the $\Omega$-particle (it consists of three s-quarks) does not allow high production 
cross sections. In fact, the $\Omega$-production is a very rare process with cross 
sections of a few \nb~only~\cite{kaidalov}. However, recent transport studies show 
that secondary scattering increases their production yield in antiproton-nucleus 
reactions~\cite{th4}. 

In conclusion, the ongoing theoretical activities, presented in this article, 
are relevant for the forthcoming experiments on in-medium superstrange hypernuclear 
physics. A more 
collaborative work between different scientific communities is required to explore 
in full detail the complex and so far unobserved regions of the nuclear and hadronic 
equation 
of state. Hypernuclear physics is a fascinating field of research, which agglutinates 
the microcosmos of nuclear and hadron physics with the macrocosmos of nuclear 
astrophysics. 

\section*{Acknowledgements}

The results in this article were obtained in collaboration with many colleagues over 
the last years. At first, I would like to thank H.H.~Wolter, C.~Fuchs, 
Th.~von~Chossy and M.~Weigel from the LMU-Munich, who introduced 
me into the world of hadron physics. I also would like to thank M.~Di~Toro, M.~Colonna, 
A.~Bonasera, J.~Rizzo and G.~Ferini for collaborations related to strangeness 
production. The collaborations with the colleagues from the experimental side were also 
very important to understand the reliability of the theoretical results. In this 
respect I would like to thank W.~Reisdorf, W.~Trautmann, Y.~Leifels, N.~Hermann, 
L.~Fabbietti, T.R.~Saito, C.~Rappold and J.~Pochodzalla. For intensive discussions 
I would like to thank the colleagues from other theoretical groups, in particular, 
A.~Botvina, S.~Typel, J.~Aichelin, C.~Hartnack, E.~Bratkovskaya, I.~Mishustin, 
P.~Danielewicz, A.~Ono, and Amand~Faessler, among many others. I would like also 
to thank my 
greek colleagues, G.A.~Lalazissis, Ch.~Moustakidis V.~Prassa from the Aristotle 
University of Thessaloniki for further helpful discussions. Finally, a great 
thank you goes to the Giessen-group. 
I would like to particularly thank U. Mosel and H. Lenske for their intensive support 
during my stay in Giessen, and W.~Cassing for deep discussions into QCD-related 
topics. I would like to thank also A.~Larionov, N.~Tsoneva, and the members of the 
GiBUU-group, in particular, O.~Buss, K.~Gallmeister and J.~Weil, for many discussions 
related to theoretical and computational issues. I would like to thank especially 
M.~Kaskulov for the great collaboration concerning the improvement of conventional 
RMF theory.

\end{document}